\documentclass[12pt,epsf]{article}

\usepackage{times}
\usepackage{graphicx}
\usepackage{amsfonts}
\usepackage{amsmath}
\usepackage{amssymb}
\usepackage{amstext}
\usepackage{amsthm}

\usepackage{epsfig}
\usepackage{color}
\psfigdriver{dvips}
\usepackage{latexsym}
\usepackage[matrix,curve]{xypic}
\usepackage{rotating}
\rotdriver{dvips}

\begin{document}

\baselineskip 6mm
\renewcommand{\thefootnote}{\fnsymbol{footnote}}


\newcommand{\nc}{\newcommand}
\newcommand{\rnc}{\renewcommand}


\rnc{\baselinestretch}{1.24}    
\setlength{\jot}{6pt}       
\rnc{\arraystretch}{1.24}   

\makeatletter
\rnc{\theequation}{\thesection.\arabic{equation}}
\@addtoreset{equation}{section}
\makeatother



\nc{\be}{\begin{equation}}

\nc{\ee}{\end{equation}}

\nc{\bea}{\begin{eqnarray}}

\nc{\eea}{\end{eqnarray}}

\nc{\xx}{\nonumber\\}

\nc{\ct}{\cite}

\nc{\la}{\label}

\nc{\eq}[1]{(\ref{#1})}

\nc{\newcaption}[1]{\centerline{\parbox{6in}{\caption{#1}}}}

\nc{\fig}[3]{

\begin{figure}
\centerline{\epsfxsize=#1\epsfbox{#2.eps}}
\newcaption{#3. \label{#2}}
\end{figure}
}


\def\CA{{\cal A}}
\def\CC{{\cal C}}
\def\CD{{\cal D}}
\def\CE{{\cal E}}
\def\CF{{\cal F}}
\def\CG{{\cal G}}
\def\CH{{\cal H}}
\def\CK{{\cal K}}
\def\CL{{\cal L}}
\def\CM{{\cal M}}
\def\CN{{\cal N}}
\def\CO{{\cal O}}
\def\CP{{\cal P}}
\def\CR{{\cal R}}
\def\CS{{\cal S}}
\def\CU{{\cal U}}
\def\CV{{\cal V}}
\def\CW{{\cal W}}
\def\CY{{\cal Y}}
\def\CZ{{\cal Z}}


\def\IB{{\hbox{{\rm I}\kern-.2em\hbox{\rm B}}}}
\def\IC{\,\,{\hbox{{\rm I}\kern-.50em\hbox{\bf C}}}}
\def\ID{{\hbox{{\rm I}\kern-.2em\hbox{\rm D}}}}
\def\IF{{\hbox{{\rm I}\kern-.2em\hbox{\rm F}}}}
\def\IH{{\hbox{{\rm I}\kern-.2em\hbox{\rm H}}}}
\def\IN{{\hbox{{\rm I}\kern-.2em\hbox{\rm N}}}}
\def\IP{{\hbox{{\rm I}\kern-.2em\hbox{\rm P}}}}
\def\IR{{\hbox{{\rm I}\kern-.2em\hbox{\rm R}}}}
\def\IZ{{\hbox{{\rm Z}\kern-.4em\hbox{\rm Z}}}}


\def\a{\alpha}
\def\b{\beta}
\def\d{\delta}
\def\ep{\epsilon}
\def\ga{\gamma}
\def\k{\kappa}
\def\l{\lambda}
\def\s{\sigma}
\def\t{\theta}
\def\w{\omega}
\def\G{\Gamma}


\def\half{\frac{1}{2}}
\def\dint#1#2{\int\limits_{#1}^{#2}}
\def\goto{\rightarrow}
\def\para{\parallel}
\def\brac#1{\langle #1 \rangle}
\def\curl{\nabla\times}
\def\div{\nabla\cdot}
\def\p{\partial}


\def\Tr{{\rm Tr}\,}
\def\det{{\rm det}}


\def\vare{\varepsilon}
\def\zbar{\bar{z}}
\def\wbar{\bar{w}}
\def\what#1{\widehat{#1}}


\def\ad{\dot{a}}
\def\bd{\dot{b}}
\def\cd{\dot{c}}
\def\dd{\dot{d}}
\def\so{SO(4)}
\def\bfr{{\bf R}}
\def\bfc{{\bf C}}
\def\bfz{{\bf Z}}

\begin{titlepage}


\hfill\parbox{3.7cm} {{\tt arXiv:1402.5134}}

\vspace{15mm}

\begin{center}
{\Large \bf Highly Effective Action from Large $N$ Gauge Fields}

\vspace{10mm}

Hyun Seok Yang \footnote{hsyang@kias.re.kr}
\\[10mm]

{\sl Center for Quantum Spacetime, Sogang University, Seoul 121-741, Korea}

\end{center}

\thispagestyle{empty}

\vskip1cm


\centerline{\bf ABSTRACT}
\vskip 4mm
\noindent

Recently John H. Schwarz put forward a conjecture that the world-volume action of a probe $D3$-brane
in an $AdS_5 \times \mathbb{S}^5$ background of type IIB superstring theory can be reinterpreted as
the highly effective action (HEA) of four-dimensional $\mathcal{N}=4$ superconformal field theory
on the Coulomb branch. We argue that the HEA can be derived from the noncommutative (NC) field theory
representation of the AdS/CFT correspondence and the Seiberg-Witten (SW) map defining a spacetime
field redefinition between ordinary and NC gauge fields. It is based only on the well-known facts
that the master fields of large $N$ matrices are higher-dimensional NC $U(1)$ gauge fields and
the SW map is a local coordinate transformation eliminating $U(1)$ gauge fields known as
the Darboux theorem in symplectic geometry.
\\



\vspace{1cm}

\today

\end{titlepage}

\renewcommand{\thefootnote}{\arabic{footnote}}
\setcounter{footnote}{0}

\section{Introduction}

Recently John H. Schwarz conjectured \cite{schwarz} that the world-volume action of a probe $p$-brane
in a maximally (or 3/4 maximal) supersymmetric spacetime containing $AdS_{p+2}$ can be reinterpreted as
the highly effective action (HEA) of a superconformal field theory in $(p+1)$-dimensions
on the Coulomb branch. The HEA is defined by taking a conformal gauge theory on the Coulomb branch
and integrating out the massive fields, thereby obtaining an effective action in terms of massless
Abelian multiplets only. Then the HEA is conjecturally identified with the world-volume action
for a probe $p$-brane in an $AdS_{p+2} \times K$ background geometry with $N$ units of flux
threading a compact space $K$. Examples considered in \cite{schwarz} are a D3-brane in $AdS_5
\times \mathbb{S}^5$, an M2-brane in $AdS_4 \times \mathbb{S}^7/\mathbb{Z}_k$, a D2-brane
in $AdS_4 \times \mathbb{CP}^3$ and an M5-brane in $AdS_7 \times \mathbb{S}^4$.
This conjecture was driven by a guiding principle \cite{schwarz}: ``Take coincidences seriously,"
with the observation that the probe brane theory has all of the expected symmetries and dualities.
The brane actions fully incorporate the symmetry of the background as an exact global symmetry
of the world-volume theory. For example, in the case of a D3-brane in $AdS_5
\times \mathbb{S}^5$, this symmetry is the superconformal group $PSU(2,2|4)$.
In this example, it also includes the $SL(2, \mathbb{Z})$ duality group, which is known to be an
exact symmetry of type IIB superstring theory. This conjecture may be further strengthened by showing
that the world-volume actions describing probe branes in AdS space exhibit
not only (super)conformal symmetry but also dual (super)conformal symmetry and, taken together,
have an infinite-dimensional Yangian-like symmetry.\footnote{Indeed this problem was addressed by
A. Lipstein and J. H. Schwarz in arXiv:1311.6067. But, unfortunately, this paper was withdrawn
due to an error in some equation.} There have also been earlier
works \cite{ads-cft1,cfs-p1,cfs-p2,cfs-p3,cfs-p4} to note the conformal symmetry of
the worldvolume theory of a $p$-brane in an AdS background as well as
works \cite{hea-col1,hea-col2,hea-col3,ferrari}
to emphasize the relationship between probe-brane actions and low-energy effective actions
on the Coulomb branch.

In this paper we will argue that the HEA can be derived from the noncommutative (NC) field theory
representation of the AdS/CFT correspondence as recently formulated in \cite{mypaper}
(see, in particular, section 6). Our argument is based only on the well-known facts that
the master fields of large $N$ matrices are higher-dimensional NC $U(1)$ gauge fields \cite{japan-matrix,nc-seiberg,hsy-epjc09,hsy-jhep09} and the Seiberg-Witten (SW) map
\cite{ncft-sw} defining a spacetime field redefinition between ordinary and NC gauge fields
is a local coordinate transformation eliminating $U(1)$ gauge fields via the Darboux theorem
in symplectic geometry \cite{cornalba,jur-sch,liu,hsy-ijmp09,hsy-jhep09}.
The underlying math for the argument is rather fundamental. For simplicity, let us consider
two-dimensional NC space, denoted by $\mathbb{R}^2_\theta$, whose coordinates obey the commutation relation
\begin{equation}\label{nc2-space}
    [y^1, y^2] = i \theta
\end{equation}
where $\theta > 0$ is a constant parameter measuring the noncommutativity of the space $\mathbb{R}^2_\theta$.
If we define annihilation and creation operators as
\begin{equation}\label{ancr}
    a = \frac{y^{1} + i y^{2}}{\sqrt{2\theta}}, \qquad
    a^\dagger = \frac{y^{1} - i y^{2}}{\sqrt{2\theta}},
\end{equation}
the NC algebra \eq{nc2-space} of $\mathbb{R}^2_\theta$ reduces to the Heisenberg algebra
of harmonic oscillator, i.e.,
\begin{equation}\label{haho}
    [a, a^\dagger]= 1.
\end{equation}
The representation space of the Heisenberg algebra (\ref{haho}) is given by the Fock space defined by
\begin{equation}\label{fock-space}
    \mathcal{H} = \{| n \rangle | \; n \in \mathbb{Z}_{\geq 0} \},
\end{equation}
which is orthonormal, i.e., $\langle n| m \rangle = \delta_{n,m}$ and
complete, i.e., $\sum_{n = 0}^{\infty} | n \rangle \langle n | = \mathbf{1}_{\mathcal{H}}$,
as is well-known from quantum mechanics.

A crucial, though elementary, fact for our argument is that the NC space $\mathbb{R}^2_\theta$
admits an infinite-dimensional separable Hilbert space \eq{fock-space} \cite{ncft-rev}.
Let us apply this elementary fact to dynamical fields defined on $\mathbb{R}^{d-1,1} \times \mathbb{R}^2_\theta$
with local coordinates $(x^\mu, y^1, y^2)$ where $\mathbb{R}^{d-1,1} \ni x^\mu$ is a $d$-dimensional
Minkowski spacetime. Consider two arbitrary fields $\widehat{\Phi}_1(x,y)$ and $\widehat{\Phi}_2(x,y)$
on $\mathbb{R}^{d-1,1} \times \mathbb{R}^2_\theta$.
In quantum mechanics physical observables are considered as operators acting on a Hilbert space.
Similarly the dynamical variables $\widehat{\Phi}_1(x,y)$ and $\widehat{\Phi}_2(x,y)$ can be regarded as
operators acting on the Hilbert space $\mathcal{H}$ which are elements of the deformed algebra $C^\infty(\mathbb{R}^{d-1,1}) \otimes \mathcal{A}_\theta$. Thus one can represent the operators acting
on the Fock space (\ref{fock-space}) as $N \times N$ matrices in $\mathrm{End}(\mathcal{H})
\equiv \mathcal{A}_N$ where $N = \mathrm{dim}(\mathcal{H}) \to \infty$:
\begin{eqnarray}\label{matrix-rep}
     && \widehat{\Phi}_1(x,y) = \sum_{n,m=0}^\infty | n \rangle \langle n| \widehat{\Phi}_1 (x, y)
     | m \rangle \langle m| := \sum_{n,m=0}^\infty (\Phi_1)_{nm} (x) | n \rangle \langle m|, \nonumber \\
     && \widehat{\Phi}_2(x,y) = \sum_{n,m=0}^\infty | n \rangle \langle n| \widehat{\Phi}_2 (x,y)
     | m \rangle \langle m| := \sum_{n,m=0}^\infty (\Phi_2)_{nm}(x) | n \rangle \langle m|,
\end{eqnarray}
where $\Phi_1 (x)$ and $\Phi_2 (x)$ are $N \times N$ matrices in $C^\infty(\mathbb{R}^{d-1,1})
\otimes \mathcal{A}_N$. Then one gets a natural composition rule for the products
\begin{eqnarray}\label{matrix-comp}
 (\widehat{\Phi}_1 \star \widehat{\Phi}_2) (x,y) &=& \sum_{n,l,m=0}^\infty | n \rangle \langle n|
 \widehat{\Phi}_1 (x,y) | l \rangle \langle l| \widehat{\Phi}_2(x,y) | m \rangle \langle m| \nonumber \\
      &=& \sum_{n,l,m=0}^\infty (\Phi_1)_{nl} (x) (\Phi_2)_{lm} (x) | n \rangle \langle m|.
\end{eqnarray}
The above composition rule implies that the ordering in the NC algebra $\mathcal{A}_\theta$
is compatible with the ordering in the matrix algebra $\mathcal{A}_N$ and so it is straightforward to
translate multiplications of NC fields in $\mathcal{A}_\theta$ into those of matrices in $\mathcal{A}_N$
using the matrix representation (\ref{matrix-rep}) without any ordering ambiguity.

It is easy to generalize the matrix representation to $2n$-dimensional NC space $\mathbb{R}^{2n}_{\theta}$
whose coordinate generators obey the commutation relation
\begin{equation}\label{extra-nc2n}
    [y^a, y^b] = i \theta^{ab}, \qquad a, b = 1, \cdots, 2n,
\end{equation}
where the Poisson bivector $\theta = \frac{1}{2} \theta^{ab} \frac{\partial}{\partial y^a} \bigwedge \frac{\partial}{\partial y^b}$ is assumed to be invertible and so $B \equiv \theta^{-1}$ defines
a symplectic structure on $\mathbb{R}^{2n}$. Consider a $D=(d+2n)$-dimensional
NC space $\mathbb{R}^{d-1,1} \times \mathbb{R}^{2n}_{\theta}$ with coordinates $Y^M = (x^\mu, y^a), \;
M = 0, 1, \cdots, D-1, \; \mu=0, 1, \cdots, d-1$. The star product for smooth functions
$\widehat{f}(Y), \widehat{g}(Y) \in C^\infty (\mathbb{R}^{D-1,1})$ is defined by
\begin{equation}\label{star-prod}
    (\widehat{f} \star \widehat{g}) (Y) = e^{\frac{i}{2} \theta^{ab} \frac{\partial}{\partial y^a}
    \otimes \frac{\partial}{\partial z^b}} \widehat{f}(x,y) \widehat{g}(x, z)|_{y=z}.
\end{equation}
Therefore, in order to formulate a gauge theory on $\mathbb{R}^{d-1,1} \times \mathbb{R}^{2n}_{\theta}$,
it is necessary to dictate the gauge covariance under the NC star product \eq{star-prod}.
The covariant field strength of NC $U(1)$ gauge fields $\widehat{A}_M (Y)
= (\widehat{A}_\mu, \widehat{A}_a)(x,y)$ is then given by
\begin{equation}\label{d-ncfs}
    \widehat{F}_{MN}(Y) = \partial_M \widehat{A}_N (Y) - \partial_N \widehat{A}_M (Y)
    - i[\widehat{A}_M, \widehat{A}_N]_\star (Y).
\end{equation}
Using the matrix representation \eq{matrix-rep},
one can show \cite{japan-matrix,nc-seiberg,hsy-epjc09,hsy-jhep09} that the $D=(d+2n)$-dimensional
NC $U(1)$ gauge theory is exactly mapped to the $d$-dimensional $U(N \to \infty)$ Yang-Mills theory:
\begin{eqnarray} \label{equiv-ncu1}
 S &=& - \frac{1}{4 G_{YM}^2} \int d^D Y (\widehat{F}_{MN} - B_{MN})^2 \\
 \label{equiv-u1un}
   &=& - \frac{1}{g_{YM}^2} \int d^d x \mathrm{Tr} \Bigl( \frac{1}{4} F_{\mu\nu}F^{\mu\nu}
   + \frac{1}{2} D_\mu \Phi_a D^\mu \Phi^a - \frac{1}{4}[\Phi_a, \Phi_b]^2 \Bigr)
\end{eqnarray}
where $G_{YM}^2 = (2 \pi )^n |\mathrm{Pf}\theta| g_{YM}^2$ and $B_{MN} = \left(
                  \begin{array}{cc}
                    0 & 0 \\
                  0 & B_{ab} \\
                  \end{array}
                \right)$.
We refer more details to the section 6.1 of Ref. \cite{mypaper}.

We emphasize that the equivalence between the $D$-dimensional NC $U(1)$ gauge theory (\ref{equiv-ncu1})
and $d$-dimensional $U(N \to \infty)$ Yang-Mill theory (\ref{equiv-u1un}) is an exact mathematical
identity, not a dimensional reduction, and has been known long ago, for example,
in \cite{japan-matrix,nc-seiberg}.
A remarkable point is that the resulting matrix models or large $N$ gauge theories described
by the action (\ref{equiv-u1un}) arise as a nonperturbative formulation of string/M theories.
For instance, we get the IKKT matrix model for $d=0$ \cite{ikkt}, the BFSS matrix quantum mechanics
for $d=1$ \cite{bfss} and the matrix string theory for $d=2$ \cite{mst}.
The most interesting case arises for $d=4$ and $n=3$ which suggests an engrossing duality that
the 10-dimensional NC $U(1)$ gauge theory on $\mathbb{R}^{3,1} \times \mathbb{R}^{6}_{\theta}$ is
equivalent to the bosonic action of 4-dimensional $\mathcal{N} = 4$ supersymmetric $U(N)$ Yang-Mills theory,
which is the large $N$ gauge theory of the AdS/CFT duality \cite{ads-cft1,ads-cft2,ads-cft3}.
According to the large $N$ duality or gauge/gravity duality, the large $N$ matrix model (\ref{equiv-u1un})
is dual to a higher dimensional gravity or string theory.
Hence it should not be surprising that the $D$-dimensional NC $U(1)$ gauge theory
should describe a theory of gravity (or a string theory) in $D$ dimensions.
Nevertheless the possibility that gravity can emerge from NC $U(1)$ gauge fields
has been largely ignored until recently. But the emergent gravity picture based
on NC $U(1)$ gauge theory \cite{mypaper,hsy-jhep09,hsy-jpcs12} debunks that this coincidence did not arise
by some fortuity. Here we want to take an advantage following the advice
of John H. Schwarz \cite{schwarz}: ``Take coincidences seriously."

In this paper, we will seriously take the equivalence between the $D$-dimensional NC $U(1)$
gauge theory (\ref{equiv-ncu1}) and $d$-dimensional $U(N \to \infty)$ Yang-Mill theory (\ref{equiv-u1un})
to derive the HEA conjectured in \cite{schwarz}. It is to be hoped that we also clarify why the emergent
gravity from NC gauge fields is actually the manifestation of the gauge/gravity duality or
large $N$ duality in string/M theories. We think that the emergent gravity from NC gauge fields opens
a lucid avenue to understand the gauge/gravity duality such as the AdS/CFT correspondence.
While the large $N$ duality is still a conjectural duality and its understanding is far from being complete
to identify an underlying first principle for the duality, it is possible \cite{mypaper,hsy-jhep09,hsy-jpcs12}
to reasonably identify the first principle for the emergent gravity from NC $U(1)$ gauge fields and
to derive in a systematic way gravitational variables from gauge theory quantities.
Moreover it can be shown \cite{mypaper} that the 4-dimensional $\mathcal{N} = 4$ supersymmetric $U(N)$
Yang-Mills theory is equivalent to the 10-dimensional $\mathcal{N} = 1$ supersymmetric
NC $U(1)$ gauge theory on $\mathbb{R}^{3,1} \times \mathbb{R}^{6}_{\theta}$ if we consider
the Moyal-Heisenberg vacuum (\ref{extra-nc2n}) which is a consistent solution of
the former -- the $\mathcal{N} = 4$ super Yang-Mills theory. Here is a foothold for our departure.

The paper is organized as follows. In section 2 we review the result in Ref. \cite{mypaper} showing
that the four-dimensional $\mathcal{N}=4$ superconformal field theory on the Coulomb branch defined
by the NC space \eq{extra-nc2n} is equivalent to the ten-dimensional $\mathcal{N}=1$
supersymmetric NC $U(1)$ gauge theory. In section 3 we consider the ten-dimensional $\mathcal{N}=1$
NC $U(1)$ super Yang-Mills theory \eq{10dsym-action} as a nontrivial leading approximation
of the supersymmetric completion of the NC DBI action. The supersymmetric completion is
postponed to section 5. In section 4, we identify a commutative DBI action which is mapped to the NC one
by the exact SW map defining a spacetime field redefinition between ordinary and NC gauge fields \cite{ncft-sw}.
It is observed that the spacetime geometry dual to four-dimensional
large $N$ matrices or ten-dimensional NC $U(1)$ gauge fields is simply derived from the Darboux transformation
eliminating $U(1)$ gauge fields whose statement is known as the Darboux theorem in symplectic geometry.
We also identify a possible candidate giving rise to $AdS_5 \times \mathbb{S}^5$ geometry.
It is shown and will also be checked in appendix A that the duality between NC $U(1)$ gauge fields and
gravitational fields is the SW map between commutative and NC $U(1)$ gauge fields. See Eq. \eq{rexp-dbi}.
We thus argue that the emergent gravity from NC gauge fields is the manifestation of the gauge/gravity
duality or large $N$ duality in string/M theories \cite{mypaper}.
In section 5, we derive the worldvolume action of
a probe D3-brane in $AdS_5 \times \mathbb{S}^5$ geometry from the DBI action of ten-dimensional NC $U(1)$
gauge fields which was obtained from the four-dimensional $\mathcal{N}=4$ superconformal field theory
on the Coulomb branch. We consider a supersymmetric D9-brane with the local $\kappa$-symmetry \cite{sdbi1,sdbi12,sdbi3,sdbi4,sdbi5,sdbi6} to yield the supersymmetric version of DBI actions.
We finally identify the supersymmetric worldvolume action of a probe D3-brane in $AdS_5 \times \mathbb{S}^5$ geometry with the HEA conjectured by John H. Schwarz \cite{schwarz}.
Our approach sheds light on why $N=1$ (i.e., Abelian gauge group) is the proper choice for the HEA
which was elusive in the original conjecture (see the discussion in section 5 of Ref. \cite{schwarz}).
In section 6, we discuss why the emergent gravity from NC gauge fields provides a lucid avenue
to understand the gauge/gravity duality such as the AdS/CFT correspondence \cite{ads-cft1,ads-cft2,ads-cft3}.
We conclude the paper with a few speculative remarks. In appendix A,
we demonstrate how to determine $2n$-dimensional K\"ahler metrics from $U(1)$ gauge fields by solving
the identities (\ref{dbi-idc}) and (\ref{dbi-idn}) between DBI actions
which are underlying equations for our argument. In particular, we show that Calabi-Yau $n$-folds
for $n=2$ and $3$ arise from symplectic $U(1)$ instantons in four and six dimensions, respectively.

\section{NC $U(1)$ gauge fields from large $N$ matrices}

The AdS/CFT correspondence \cite{ads-cft1,ads-cft2,ads-cft3} implies that a wide variety of quantum field
theories provide a nonperturbative realization of quantum gravity. In the AdS/CFT duality,
the dynamical variables are large $N$ matrices and so gravitational physics at a fundamental level
is described by NC operators. We argued in \cite{mypaper} that the AdS/CFT correspondence is a particular
case of emergent gravity from NC U(1) gauge fields. An underlying argumentation is to realize
the equivalence between the actions \eq{equiv-ncu1} and \eq{equiv-u1un} in a reverse way by observing that
the Moyal-Heisenberg vacuum (\ref{extra-nc2n}) is a consistent vacuum solution of
the $\mathcal{N} = 4$ super Yang-Mills theory.

It is easy to understand an underlying logic and so we recapitulate only the essential points deferring
to \cite{mypaper} on a detailed description.
The action of four-dimensional $\mathcal{N} = 4$ super Yang-Mills theory is given by \cite{n=4sym}
\begin{eqnarray}\label{n=4-action}
    S &=& \int d^4 x \mathrm{Tr} \left\{- \frac{1}{4} F_{\mu\nu} F^{\mu\nu} - \frac{1}{2} D_\mu \Phi_a
    D^\mu \Phi_a + \frac{g^2}{4}[\Phi_a, \Phi_b]^2 + i \overline{\lambda}_i
    \overline{\sigma}^\mu D_\mu \lambda^i \right. \nonumber \\
    && \qquad \qquad \left. + \frac{ig}{2} \overline{\Sigma}^a_{ij} \lambda^i [ \Phi_a, \lambda^j]
    - \frac{ig}{2} \Sigma^{a, ij} \overline{\lambda}_i [ \Phi_a, \overline{\lambda}_j] \right\}.
\end{eqnarray}
Consider a vacuum configuration defined by
\begin{equation}\label{n=4vacuum}
    \langle \Phi_a \rangle_{\mathrm{vac}} = p_a, \quad
    \langle A_\mu \rangle_{\mathrm{vac}} = 0, \quad \langle \lambda^i \rangle_{\mathrm{vac}} = 0.
\end{equation}
Assume that the vacuum expectation value (vev) $p_a \in \mathcal{A}_N \;
(N \to \infty)$ satisfies the Moyal-Heisenberg algebra
\begin{equation}\label{n=4moyal}
    [p_a, p_b] = - i B_{ab} I_{N \times N}.
\end{equation}
Of course the commutation relation (\ref{n=4moyal}) is meaningful only when we take
the limit $N \to \infty$. It is obvious that the vacuum configuration (\ref{n=4vacuum}) in this limit
is definitely a solution of the theory. We emphasize that the vev (\ref{n=4vacuum})
of adjoint scalar fields does not break four-dimensional Lorentz symmetry.
Actually the vacuum algebra (\ref{n=4moyal}) refers to NC space $\mathbb{R}_\theta^6$ if we define
$p_a \equiv  B_{ab} y^b$ and $B \equiv \theta^{-1}$.
Now fluctuations of large $N$ matrices around the vacuum (\ref{n=4vacuum}) are parameterized by
\begin{eqnarray} \label{n=4bfluct}
&& \widehat{D}_\mu (x,y) = \partial_\mu - i\widehat{A}_\mu(x,y),  \quad
\widehat{D}_a (x,y) \equiv -i\widehat{\Phi}_a (x,y) = -i\bigl( p_a + \widehat{A}_a(x,y) \bigr), \\
\label{n=4ffluct}
&& \widehat{\Psi} (x, y) = \left(
                                            \begin{array}{c}
                                              P_+ \widehat{\lambda}^i \\
                                              P_- \widetilde{\widehat{\lambda}}_i  \\
                                            \end{array}
                                          \right) (x, y),
\end{eqnarray}
where we assumed that fluctuations also depend on vacuum moduli $y^a$.
Note that, if we apply the matrix representation (\ref{matrix-rep}) to the fluctuations
in Eqs. (\ref{n=4bfluct}) and (\ref{n=4ffluct}) again, we recover the original
large $N$ gauge fields in the action (\ref{n=4-action}).
Therefore let us introduce 10-dimensional coordinates $Y^M = (x^\mu, y^a), \; M = 0, 1, \cdots, 9$
and 10-dimensional connections defined by
\begin{equation}\label{10d-conn}
    \widehat{D}_M(Y) = \partial_M - i\widehat{A}_M (x,y) = (\widehat{D}_\mu, \widehat{D}_a) (x,y)
\end{equation}
whose field strength is given by
\begin{equation}\label{10d-fs}
    \widehat{F}_{MN}(Y) = i [\widehat{D}_M, \widehat{D}_N]_\star
    = \partial_M \widehat{A}_N - \partial_N \widehat{A}_M - i[\widehat{A}_M, \widehat{A}_N]_\star.
\end{equation}
Thus the correspondence between the NC $\star$-algebra $\mathcal{A}_\theta$ and
the matrix algebra $\mathcal{A}_N = \mathrm{End}(\mathcal{H})$ under the Moyal-Heisenberg
vacuum (\ref{n=4moyal}) implies that the master fields of large $N$ matrices
are higher-dimensional NC $U(1)$ gauge fields. In the end large $N$ matrices in $\mathcal{N}=4$
vector multiplet on $\mathbb{R}^{3,1}$ are mapped to NC gauge fields and their superpartners
in $\mathcal{N}=1$ vector multiplet on $\mathbb{R}^{3,1} \times \mathbb{R}_{\theta}^6$
where $\mathbb{R}_\theta^{6}$ is an extra NC space whose coordinate generators $y^a \in \mathcal{A}_\theta$
obey the commutation relation (\ref{extra-nc2n}).

Using the ordering (\ref{matrix-comp}) for $U(N)$ and NC $U(1)$ gauge fields, it is straightforward
to organize the 4-dimensional $\mathcal{N}=4 \; U(N)$ super Yang-Mills theory (\ref{n=4-action})
into the 10-dimensional $\mathcal{N}=1$ NC $U(1)$ super Yang-Mills theory with the action \cite{mypaper}
\begin{equation}\label{10dsym-action}
    S = \int d^{10} Y \left\{ - \frac{1}{4G_{YM}^2} (\widehat{F}_{MN} - B_{MN})^2
    + \frac{i}{2} \overline{\widehat{\Psi}} \Gamma^M \widehat{D}_M \widehat{\Psi} \right\}
\end{equation}
where $B$-fields take the same form as Eq. (\ref{equiv-ncu1}). Now the fermion $\widehat{\Psi}(Y)$
is a 10-dimensional gaugino, the superpartner of the 10-dimensional NC $U(1)$ gauge field $\widehat{A}_M(x)$,
that is the Majorana-Weyl spinor of $SO(9,1)$. The action (\ref{10dsym-action}) is invariant
under $\mathcal{N}=1$ supersymmetry transformations given by
\begin{equation}\label{10dsusytr}
    \delta \widehat{A}_M = i \overline{\alpha} \Gamma_M \widehat{\Psi}, \qquad
    \delta \widehat{\Psi} = \frac{1}{2} (\widehat{F}_{MN} - B_{MN}) \Gamma^{MN} \alpha.
\end{equation}
It should be remarked that the relationship between the 4-dimensional $U(N)$ super Yang-Mills
theory (\ref{n=4-action}) and 10-dimensional NC $U(1)$ super Yang-Mills theory (\ref{10dsym-action})
is not a dimensional reduction but they are exactly equivalent to each other.
Therefore any quantity in lower-dimensional $U(N)$ gauge theory can be transformed into an object
in higher-dimensional NC $U(1)$ gauge theory using the compatible ordering (\ref{matrix-comp}) \cite{mypaper}.

The coherent condensate (\ref{n=4vacuum}) is described by vev's of adjoint scalar fields.
Thus we will call the vacuum (\ref{n=4vacuum}) a ``Coulomb branch" although $[\Phi_a, \Phi_b]|_{\mathrm{vac}}
\neq 0$.\footnote{\label{n=1}The usual Coulomb branch is defined by $[\Phi_a, \Phi_b]|_{\mathrm{vac}} = 0$
and so $\langle \Phi_a \rangle_{\mathrm{vac}} = \mathrm{diag}(\alpha_{a_1}, \cdots, \alpha_{a_N})$.
In this case the gauge group $U(N)$ or $SU(N+1)$ is broken to $U(1)^N$.
But we remark that the HEA is conjectured to correspond to the choice, $N = 1$ \cite{schwarz}
while the probe brane approximation requires $N \to \infty$.
Therefore the conventional choice of vacuum finds difficulty in explaining
why $N=1$ (i.e., Abelian gauge group) is the proper choice for the HEA.
We emphasize that the Coulomb branch as the NC space (\ref{n=4vacuum}) is a key origin
of emergent gravity and is completely consistent with the HEA because it requires
the $N \to \infty$ limit and preserves only the $U(1)$ gauge group.
Hence our approach sheds light on why HEA preserves only the $U(1)$ gauge symmetry
in spite of $N \to \infty$ which was elusive in the original conjecture
as discussed in section 5 of Ref. \cite{schwarz}.}
However note that $[\Phi_a, \Phi_b]|_{\mathrm{vac}} = - i B_{ab} I_{N \times N}$ take values
in a center of the gauge group $U(N)$, which may be identified with the unbroken $U(1)$ gauge group.
Hence the Coulombic vacuum (\ref{n=4vacuum}) is compatible with the usual definition of the Coulomb branch.
We also remark that the conformal symmetry of 4-dimensional $\mathcal{N} = 4$ super Yang-Mills theory
is spontaneously broken by the vev (\ref{n=4vacuum}) of scalar fields because
it introduces a NC scale $|\theta| \equiv l^2_{NC}$. But it needs not be specified because the theories
with different $\theta$'s are SW-equivalent \cite{ncft-sw}. These are also a typical feature of
the Coulomb branch.

Under a Coulomb branch described by the coherent condensate (\ref{n=4vacuum}),
large $N$ matrices in $\mathcal{N}=4$ supersymmetric
gauge theory can be regarded as a linear representation of operators acting on a separable Hilbert
space $\mathcal{H}$ that is the Fock space of the Moyal-Heisenberg vacuum (\ref{n=4moyal}).
Therefore an important point is that a large $N$ matrix $\Phi(x)$ on four-dimensional spacetime $\mathbb{R}^{3,1}$ in the limit $N \to \infty$ on the Coulomb branch (\ref{n=4vacuum}) can be represented by its master field $\widehat{\Phi}(x,y)$ which is a higher-dimensional NC $U(1)$ gauge field or its superpartner.
Since the large $N$ gauge theory (\ref{n=4-action}) on the Coulomb branch (\ref{n=4vacuum}) is mathematically equivalent to the NC $U(1)$ gauge theory described by the action (\ref{10dsym-action}), it should be possible to isomorphically map the 10-dimensional NC $U(1)$ super Yang-Mills theory to a 10-dimensional type IIB supergravity according to the AdS/CFT correspondence \cite{ads-cft1,ads-cft2,ads-cft3}.
Indeed the emergent gravity from NC $U(1)$ gauge fields provides the first principle to found
the large $N$ duality or gauge/gravity duality in a systematic way \cite{mypaper,hsy-jhep09,hsy-jpcs12}.

\section{Commutative and NC D-branes}

The worldvolume action for a D$p$-brane can be viewed as $(p+1)$-dimensional nonlinear sigma model
with a target space $M$ where
the embedding functions $X^M(\sigma)$ define a map $X: W \to M$ from the $(p+1)$-dimensional
worldvolume $W$ with coordinates $\sigma^\alpha \; (\alpha = 0, 1, \cdots, p)$ to the target space $M$
with coordinates $X^M \; (M = 0, 1, \cdots, 9)$. This embedding induces a worldvolume metric
\begin{equation}\label{ind-metric}
    h_{\alpha\beta} = g_{MN} (X) \partial_\alpha X^M \partial_\beta X^N.
\end{equation}
The D-brane action in general contains a dilaton coupling $e^{-\phi}$ where $\phi$ is the 10-dimensional
dilaton field. Then the string coupling constant is defined by $g_s =e^{\langle \phi \rangle}$ where
the vev $\langle \phi \rangle$ at hand is assumed to be constant.
The worldvolume also carries $U(1)$ gauge fields $A_\alpha (\sigma)$ with field strength
\begin{equation}\label{wvgauge}
    F_{\alpha\beta} = \partial_\alpha A_\beta - \partial_\beta A_\alpha.
\end{equation}
Recall that the Dirac-Born-Infeld (DBI) action is a nonlinear generalization of electrodynamics
with self-interactions of $U(1)$ gauge fields and reproduces the usual Maxwell theory at quadratic order.
In string theory a generalization of this action appears in the context of D$p$-branes.
Open strings ending on the D$p$-brane couple directly to closed string background fields
$(g_{MN}, B_{MN}, \phi)$ in the bulk.
A low energy effective field theory deduced from the open string dynamics on a single D-brane
is obtained by integrating out all the massive modes, keeping only massless fields which are slowly varying
at the string scale $\kappa \equiv 2 \pi \alpha'$. The DBI action describes the dynamics of
$U(1)$ gauge fields on a D-brane worldvolume in the approximation of slowly varying fields,
$\sqrt{\kappa} |\frac{\partial F}{F}| \ll 1$, in the sense keeping field strengths
(without restriction on their size) but not their derivatives.
The resulting DBI action on a D$p$-brane is given by
\begin{equation}\label{cdbi}
    S_1 = - T_{\mathrm{D}p} \int_W d^{p+1} \sigma \sqrt{-\det \bigl(h
    + \kappa \mathcal{F} \bigr)} + \mathcal{O} (\sqrt{\kappa} \partial F, \cdots),
\end{equation}
where
\begin{equation}\label{total-f}
\mathcal{F} \equiv B + F
\end{equation}
is the total $U(1)$ field strength and the D$p$-brane tension is given by
\begin{equation}\label{cdp-tension}
    T_{\mathrm{D}p} = \frac{2 \pi}{g_s (2 \pi \kappa)^{\frac{p+1}{2}}}.
\end{equation}
In general the DBI action (\ref{cdbi}) contains derivative corrections
$\mathcal{O} (\sqrt{\kappa} \partial F, \cdots)$. However we will ignore possible terms involving
higher derivatives of fields since we are mostly interested in the approximation that worldvolume
fields are slowly varying. We will also consider the probe brane approximation ignoring
the backreaction of the brane on the geometry and the other background fields.
The worldvolume theory of a D-brane is given as the sum of two terms $S = S_1 + S_2$.
The first term $S_1$ is given by the DBI action (\ref{cdbi}) and the second term $S_2$ is
the form of the Wess-Zumino-type given by
\begin{equation}\label{wz-action}
    S_2 = \int_W C_{RR} \wedge e^{\kappa \mathcal{F}}
\end{equation}
where the coupling to background RR $n$-form gauge fields is collected in the formal sum
\begin{equation}\label{rr-field}
    C_{RR} = \bigoplus_{n=0}^{10} C_n.
\end{equation}
The coupling $S_2$ is a characteristic feature of D-branes that they carry an RR charge \cite{polchinski}
and support the worldvolume gauge fields \eq{wvgauge}.

Some important remarks are in order. The DBI action (\ref{cdbi}) respects several local gauge symmetries.
It has $(p+1)$-dimensional general coordinate invariance since the integrand transforms
as a scalar density in Diff$(W)$. It also admits the so-called $\Lambda$-symmetry:
\begin{equation}\label{l-symmetry}
    (B, A) \mapsto (B-d\Lambda, A + \Lambda)
\end{equation}
where the two-form $B \equiv X^* \bigl( B_{\mathrm{bulk}} \bigr)$ is the pull-back of target space
$B$-field $B_{\mathrm{bulk}}$ to the worldvolume $W$ and the gauge parameter $\Lambda$ is a one-form
in $\Gamma(T^* W)$. Let $(W, B)$ be a symplectic manifold. The symplectic structure $B$ is a nondegenerate,
closed two-form, i.e. $dB=0$, and so it can be locally written as $B = d \xi$ by the Poincar\'e lemma.
The $B$-field transformation (\ref{l-symmetry}) can then be understood as a shift of the canonical
one-form, $\xi \to \xi - \Lambda$. An important point for us is that the symplectic structure defines
a bundle isomorphism $B: TW \to T^* W$ by $X \mapsto \Lambda = - \iota_X B$.
Thus the $B$-field transformation (\ref{l-symmetry}) is equivalent to
$(B, A) \mapsto (B + \mathcal{L}_X B, A - \iota_X B)$ where $\mathcal{L}_X = d\iota_X
+ \iota_X d$ is the Lie derivative with respect to the vector field $X$. Since vector fields
are infinitesimal generators of local coordinate transformations, in other words,
Lie algebra generators of Diff$(W)$, the $B$-field transformation (\ref{l-symmetry}) can be identified
with a coordinate transformation generated by a vector field $X \in \Gamma(TW)$.
Consequently the $\Lambda$-symmetry (\ref{l-symmetry}) can be considered on par
with diffeomorphisms \cite{mypaper,hsy-jhep09}. Moreover it is well-known \cite{sdbi1,sdbi12,sdbi3,sdbi4,sdbi5,sdbi6}
that the D-brane worldvolume theory has a local fermionic symmetry called ``$\kappa$-symmetry"
if fermion coordinates $\psi^\alpha \; (\alpha=1, \cdots, 32)$
are included in the target spacetime with supercoordinates $Z^{\mathbf{M}} = (X^M, \psi^\alpha)$.
See a recent review \cite{jsimon} for brane effective actions with the $\kappa$-symmetry.
In sum, the worldvolume theory of a supersymmetric D-brane admits the following local gauge symmetries:
(I) Diff$(W)$, (II) $\Lambda$-symmetry, and (III) $\kappa$-symmetry.

We can use the general coordinate invariance of the action $S = S_1 + S_2$ to eliminate unphysical
degrees of freedom. We choose a static gauge so that $X^M = \bigl( x^\mu (\sigma), \phi^a (\sigma) \bigr)
= \bigl(\delta^\mu_\alpha \sigma^\alpha, \phi^a(x) \bigr)$ where $\mu = 0, \cdots, p$ and
$a = p+1, \cdots, 9$. The $(9-p)$ coordinates $\phi^a (x)$ will be identified as the worldvolume scalar
fields of the D$p$-brane. In this gauge the metric \eq{ind-metric} becomes
\begin{equation}\label{wv-metric}
    h_{\mu\nu} = \eta_{\mu\nu} + \partial_\mu \phi^a \partial_\nu \phi^a
\end{equation}
where we assumed $g_{MN} (X) = \eta_{MN}$ for the target spacetime.
Now we focus on a D9-brane for which there are no worldvolume scalar fields, i.e., $\phi^a = 0$
and so $h_{MN} = g_{MN}$. Suppose that the D9-brane supports the two-form $B$-field with $\mathrm{rank}(B) = 6$.
In this case it is convenient to split the worldvolume coordinates $X^M = \sigma^M$ in the static gauge
into two parts, $X^M = (x^\mu, z^a), \; \mu=0,1,2, 3, \; a=1, \cdots, 6$,
so that $B = \frac{1}{2} B_{ab} dz^a \wedge dz^b$.
Then the total field strength \eq{total-f} takes the form
\begin{equation}\label{matrix-10f}
\mathcal{F}_{MN} = \left(
  \begin{array}{cc}
    F_{\mu\nu} & F_{\mu a} \\
    F_{a \mu} & B_{ab} + F_{ab} \\
  \end{array}
\right).
\end{equation}
It is well-known \cite{ncft-sw} that the open string gives rise to the NC geometry
when the two-form $B$-field is present on a D-brane worldvolume.
The D-brane dynamics in the static gauge is then described by $U(1)$ gauge fields on
a NC spacetime with coordinates $Y^M = (x^\mu, y^a)$ obeying the commutation relation \eq{extra-nc2n}.
The resulting DBI action on the NC D9-brane is given by
\begin{equation}\label{ncdbi}
    \widehat{S}_1 = - T_{9} \int d^{10} Y \sqrt{-\det \bigl(G
    + \kappa (\widehat{F} + \Phi ) \bigr)} + \mathcal{O} (\sqrt{\kappa} \widehat{D}\widehat{F}, \cdots),
\end{equation}
where the NC $U(1)$ field strength $\widehat{F}_{MN}(Y)$ is given by Eq. \eq{d-ncfs} and
the NC D9-brane tension is
\begin{equation}\label{dp-tension}
    T_{9} = \frac{2 \pi}{G_s (2 \pi \kappa)^{5}}.
\end{equation}
The open string moduli $(G, \Phi, G_s)$ in the NC description \eq{ncdbi} are related
to the closed string moduli $(g, B, g_s)$ in the commutative description \eq{cdbi} by \cite{ncft-sw}
\begin{eqnarray}\label{open-closed1}
    && \frac{1}{g + \kappa B} = \frac{1}{G + \kappa \Phi} + \frac{\theta}{\kappa}, \\
    \label{open-closed2}
    && G_s = g_s \sqrt{\frac{\det(G + \kappa \Phi)}{\det (g + \kappa B)}}
    = g_s \left( \frac{\det G}{\det g} \right)^{\frac{1}{4}},
\end{eqnarray}
where the two-form $\Phi$ parameterizes some freedom in the description of commutative and
NC gauge theories. It is worthwhile to remark that the NC DBI action \eq{ncdbi} can be obtained
by applying the (exact) SW map to the commutative one \eq{cdbi} \cite{liu,jsw2,ban-yan},
as will be shown later. Similarly the Wess-Zumino-type term $\widehat{S}_2$ for the NC D9-brane
can be obtained from the RR couplings in Eq. \eq{wz-action} for a commutative D9-brane
by considering the (exact) SW map \cite{liu,nc-coupling}.

Let us expand the NC DBI action \eq{ncdbi} in powers of $\kappa$.
First note that
\begin{eqnarray}\label{det-exp}
 \sqrt{-\det \bigl(G    + \kappa (\widehat{F} + \Phi ) \bigr)} &=& \sqrt{-\det G}
 \sqrt{\det (1 + \kappa M)}  \\
 &=& \sqrt{-\det G} \Bigl( 1 - \frac{\kappa^2}{4} \mathrm{Tr} M^2
 - \frac{\kappa^4}{8} \mathrm{Tr} M^4 + \frac{\kappa^4}{32} \bigl (\mathrm{Tr} M^2 \bigr)^2
 + \cdots \Bigr), \nonumber
\end{eqnarray}
where
\begin{equation}\label{def-m}
    {M_N}^Q \equiv (\widehat{F} + \Phi)_{NP}G^{PQ}
\end{equation}
and so $\mathrm{Tr} M = 0$. At nontrivial leading orders, we find
\begin{equation}\label{exp-ncym}
     \widehat{S}_1  = - T_{9} \int d^{10} Y \sqrt{-\det G} - \frac{1}{4G_{YM}^2} \int d^{10} Y \sqrt{-\det G}
     G^{MP} G^{NQ} (\widehat{F} + \Phi)_{MN}(\widehat{F} + \Phi)_{PQ} + \mathcal{O} (\kappa^4),
\end{equation}
where the 10-dimensional Yang-Mills coupling constant is given by
\begin{equation}\label{10ymcs}
    G_{YM}^2 = (\kappa^2 T_9)^{-1} = (2\pi)^4 \kappa^3 G_s.
\end{equation}
In our case at hand, the open string metric can be set to be flat, i.e., $G_{MN} = \eta_{MN}$.
The first term of $\widehat{S}_1$ is a vacuum energy due to the D-brane tension which will be
canceled against a contribution from $\widehat{S}_2$ \cite{schwarz,ads-cft1}.
The second term is precisely equal to the bosonic part of the action \eq{10dsym-action}
when the background independent prescription is employed, i.e., $\Phi = - B$ \cite{ncft-sw}.
Therefore we will consider the 10-dimensional $\mathcal{N}=1$ NC $U(1)$ super Yang-Mills theory
\eq{10dsym-action} as a nontrivial leading approximation of the supersymmetric completion of the NC DBI
action \eq{ncdbi}. The supersymmetric completion with the $\kappa$-symmetry will be discussed in section 5.

\section{AdS/CFT correspondence from NC $U(1)$ gauge fields}

In their famous paper \cite{ncft-sw}, Seiberg and Witten showed that there exists an equivalent commutative
description of the low energy effective theory for the open string ending on a NC D-brane.
From the point of view of open string sigma model, an explicit form of the effective action depends
on the regularization scheme of two-dimensional field theory.
The difference due to different regularizations is always in a choice of contact terms,
leading to the redefinition of coupling constants which are spacetime fields.
So low energy field theories defined with different regularizations should be related to each other
by the field redefinitions in spacetime. Now we will explain how the NC DBI action \eq{ncdbi} arises
from a low energy effective action in a curved background that will be identified with the HEA speculated
by John H. Schwarz \cite{schwarz}. First we identify a commutative description that is SW-equivalent
to the NC DBI action \eq{ncdbi}. From a conventional approach, the answer is obvious. It is given by
the D9-brane action \eq{cdbi} (with $p=9$) with the field strength \eq{matrix-10f}.
But, for our purpose, it is more proper to consider the NC DBI action \eq{ncdbi} as a particular
commutative limit of the full NC D9-brane described by the star product
\begin{equation}\label{10d-star-prod}
    (\widehat{f} \star \widehat{g}) (Y) = e^{\frac{i}{2} \Theta^{MN} \frac{\partial}{\partial Y^M}
    \otimes \frac{\partial}{\partial Z^N}} \widehat{f}(Y) \widehat{g}(Z)|_{Y=Z}
\end{equation}
for $\widehat{f}(Y), \widehat{g}(Y) \in C^\infty (\mathbb{R}^{10})$. We implicitly assumed
the Wick rotation, $\mathbb{R}^{9,1} \to \mathbb{R}^{10}$, although it is simply formal because
we eventually come back to the space $\mathbb{R}^{3,1} \times \mathbb{R}^{6}_\theta$.
For this purpose, it is convenient to take the split $\Theta^{MN} = (\zeta^{\mu\nu}, \theta^{ab})$
where an $SO(10)$ rotation was used to put $\zeta^{\mu a} = 0$.
We intend to understand the star product \eq{star-prod} as a particular case of
Eq. \eq{10d-star-prod} with $\zeta^{\mu\nu} = 0$.
Later we will explain why the star product \eq{10d-star-prod} is more relevant for our context,
especially, from the viewpoint of emergent spacetime. Hence we need to identify a commutative DBI action
that is SW-equivalent to the NC DBI action \eq{ncdbi}, instead, using the star product \eq{10d-star-prod}.
It is given by the D9-brane action \eq{cdbi} with the $U(1)$ field strength
\begin{equation}\label{10t-fs}
    \mathcal{F} = \frac{1}{2} \mathcal{F}_{MN} (X) dX^M \wedge dX^N = \frac{1}{2}
    \bigl(B_{MN} + F_{MN}(X) \bigr) dX^M \wedge dX^N = B + F
\end{equation}
where $B = \Theta^{-1}$ and $\mathrm{rank}(B) = 10$.
We will assume that $\mathcal{F}$ is also nondegenerate, i.e., $\det(1 + F \Theta) \neq 0$.

In order to derive the HEA, it is enough only to employ the logic expounded
in the appendix A in Ref. \cite{mypaper}. Note that $\mathcal{F}$ in Eq. \eq{10t-fs} is the gauge
invariant quantity under the $\Lambda$-symmetry \eq{l-symmetry}.
In other words, the dynamical $U(1)$ gauge fields should appear only
as the combination \eq{10t-fs}. In particular, we can use the $\Lambda$-symmetry \eq{l-symmetry} so that
the $B$-field in Eq. \eq{10t-fs} is constant.
Then $dB=0$ trivially and $B$ is nondegenerate because of $\mathrm{rank}(B) = 10$.
Therefore $(\mathbb{R}^{10}, B)$ is a symplectic manifold.
Moreover, $(\mathbb{R}^{10}, \mathcal{F})$ is also a symplectic manifold since $d\mathcal{F}=0$ and
$\mathcal{F}$ is nondegenerate by our assumption. Then we can realize an important identity
\begin{equation}\label{darboux}
  \mathcal{F} = (1 + \mathcal{L}_X) B
\end{equation}
as we explained below Eq. \eq{l-symmetry}. It implies that there exists a local coordinate transformation
$\phi \in \mathrm{Diff}(M)$ such that $\phi^* (\mathcal{F}) = B$, i.e.,
$\phi^* = (1 + \mathcal{L}_X)^{-1} \approx e^{-\mathcal{L}_X}$.
This statement is the famous theorem in symplectic geometry known as
the Darboux theorem \cite{sg-book1,sg-book2}. Its global statement is known as the Moser lemma \cite{moser}.
The Darboux theorem states that it is always possible to find a local coordinate
transformation $\phi \in \mathrm{Diff}(M)$ which eliminates dynamical $U(1)$ gauge fields
in $\mathcal{F}$. That is, in terms of local coordinates, there exists $\phi: Y \mapsto X = X(Y)$ so that
\begin{equation}\label{darboux-local}
 \bigl(B_{MN} + F_{MN}(X) \bigr) \frac{\partial X^M}{\partial Y^P}
 \frac{\partial X^N}{\partial Y^Q} = B_{PQ}.
\end{equation}
If we represent the local coordinate transformation by
\begin{equation}\label{cov-cod}
    X^M (Y) = Y^M + \Theta^{MN} \widehat{A}_N (Y),
\end{equation}
Eq. \eq{darboux-local} can be written as
\begin{equation}\label{sym-gauge}
    \mathfrak{P}^{MN} (X) \equiv \bigl(\mathcal{F}^{-1} \bigr)^{MN} (X) = \{ X^M (Y), X^N(Y) \}_\Theta
\end{equation}
where we introduced the Poisson bracket defined by
\begin{equation}\label{poisson-bra}
    \{f(Y), g(Y)\}_\Theta = \Theta^{MN} \frac{\partial f(Y)}{\partial Y^M}
 \frac{\partial g(Y)}{\partial Y^N}
\end{equation}
for $f, g  \in C^\infty (\mathbb{R}^{10})$. We will call $\widehat{A}_M (Y)$ in Eq. \eq{cov-cod}
symplectic gauge fields and  $X^M(Y)$ covariant (dynamical) coordinates.
The field strength of symplectic gauge fields is defined by
\begin{equation}\label{symp-f}
    \widehat{F}_{MN} = \partial_M \widehat{A}_N - \partial_N \widehat{A}_M
    + \{ \widehat{A}_M, \widehat{A}_N \}_\Theta.
\end{equation}
Then Eq. \eq{sym-gauge} gives us the relation
\begin{equation}\label{new-poisson}
     \mathfrak{P}^{MN} = [\Theta (B - \widehat{F})\Theta]^{MN}.
\end{equation}
By solving this equation, we yield the semi-classical version of the SW map \cite{cornalba,jur-sch,liu}:
\begin{eqnarray}\label{sw-mapf}
    && \widehat{F}_{MN} (Y) = \left( \frac{1}{1 + F\Theta} F \right)_{MN} (X), \\
    \label{sw-mapv}
    &&  d^{10} Y =  d^{10} X \sqrt{\det(1 + F\Theta)},
\end{eqnarray}
where the second equation is derived from Eq. \eq{darboux-local} by taking the determinant
on both sides.

The coordinate transformation \eq{darboux-local} leads to the identity
\begin{equation}\label{tr-dbi}
    g_{MN} + \kappa \mathcal{F}_{MN} = \bigl(\mathcal{G}_{PQ} + \kappa B_{PQ} \bigr)
    \frac{\partial Y^P}{\partial X^M} \frac{\partial Y^Q}{\partial X^N}
\end{equation}
where the dynamical (emergent) metric is defined by
\begin{equation}\label{de-metric}
 \mathcal{G}_{MN} = g_{PQ} \frac{\partial X^P}{\partial Y^M}
 \frac{\partial X^Q}{\partial Y^N}.
\end{equation}
The identity \eq{tr-dbi} in turn leads to a remarkable identity between DBI actions:
\begin{eqnarray} \label{dbi-idc}
\frac{1}{g_s} \int d^{10} X \sqrt{\det \bigl( g + \kappa \mathcal{F} \bigr)}
&=& \frac{1}{g_s} \int d^{10} Y \sqrt{\det \bigl(\mathcal{G}  + \kappa B \bigr)} \\
 \label{dbi-idn}
&=& \frac{1}{G_s} \int d^{10} Y \sqrt{\det \bigl(G  + \kappa (\widehat{F} + \Phi ) \bigr)}.
\end{eqnarray}
It is straightforward to derive the second identity \eq{dbi-idn} by using Eqs. \eq{open-closed1}
and \eq{open-closed2} and the SW maps \eq{sw-mapf} and \eq{sw-mapv}.
For the derivation of Eq. \eq{dbi-idn},
see Eq. (5.10) in Ref. \cite{liu} and section 3.4 of Ref. \cite{jsw2}.
It may be instructive to check Eq. \eq{dbi-idn} by expanding the right-hand side (RHS) of Eq. \eq{dbi-idc}
around the background $B$-field, i.e.,
\begin{eqnarray}\label{exp-dbi}
\sqrt{\det \bigl(\mathcal{G}  + \kappa B \bigr)} &=& \sqrt{\det \bigl( \kappa B \bigr)}
\sqrt{\det \Bigl(1  + \frac{M}{\kappa} \Bigr)} \nonumber \\
&=& \sqrt{\det \bigl( \kappa B \bigr)} \Bigl( 1 - \frac{1}{4\kappa^2} \mathrm{Tr} M^2
 - \frac{1}{8\kappa^4} \mathrm{Tr} M^4 + \frac{1}{32\kappa^4} \bigl (\mathrm{Tr} M^2 \bigr)^2
 + \cdots \Bigr),
\end{eqnarray}
where
\begin{equation}\label{matrix-gm}
    {M_N}^Q = \mathcal{G}_{NP} \Theta^{PQ}
\end{equation}
and
\begin{equation}\label{tr-pm}
   \mathrm{Tr} M^2 = \mathrm{Tr} (g \mathfrak{P})^2, \qquad \mathrm{Tr} M^4 = \mathrm{Tr} (g \mathfrak{P})^4.
\end{equation}
But it is not difficult to show that $\mathrm{Tr} M^{2n} = \mathrm{Tr} (g \mathfrak{P})^{2n}, \;
\mathrm{Tr} M^{2n+1} = \mathrm{Tr} (g \mathfrak{P})^{2n+1} = 0$ for $n \in \mathbb{N}$ and thus
\begin{equation}\label{det-id}
 \det \Bigl(1  + \frac{M}{\kappa} \Bigr) = \det \Bigl(1  + \frac{1}{\kappa} g \mathfrak{P} \Bigr)
\end{equation}
using the expansion of the determinant (see Eq. (4.30) in Ref. \cite{sdbi4}).
Then, using the result \eq{new-poisson}, the expansion in Eq. \eq{exp-dbi} can be arranged into the form
\begin{eqnarray}\label{rexp-dbi}
\sqrt{\det \bigl(\mathcal{G}  + \kappa B \bigr)} &=& \sqrt{\frac{\det \bigl( \kappa B \bigr)}{\det G}}
\sqrt{\det \bigl(G  + \kappa (\widehat{F} - B ) \bigr)} \nonumber \\
&=& \frac{g_s}{G_s} \sqrt{\det \bigl(G  + \kappa (\widehat{F} - B ) \bigr)},
\end{eqnarray}
where
\begin{equation}\label{metric-coupling}
    G_{MN} = - \kappa^2 (B g^{-1} B)_{MN}, \qquad G_s = g_s \sqrt{\det \bigl(\kappa B g^{-1} \bigr)}
\end{equation}
are the open string metric and coupling constant, respectively,
in the background independent prescription, i.e., $\Phi = - B$ \cite{ncft-sw}.
In order to demonstrate how $2n$-dimensional K\"ahler metrics arise from $U(1)$ gauge fields,
in appendix A, we will solve the identities (\ref{dbi-idc}) and (\ref{dbi-idn}).
In particular, it is shown that Calabi-Yau $n$-folds for $n=2$ and $3$ are emergent
from symplectic $U(1)$ instantons in four and six dimensions, respectively.

NC $U(1)$ gauge fields are obtained by quantizing symplectic gauge fields.
The quantization in our case is simply defined by the canonical quantization of
the Poisson algebra $\mathfrak{P} = (C^\infty(\mathbb{R}^{10}), \{-,-\}_\Theta)$.
The quantization map $\mathcal{Q}: C^\infty(\mathbb{R}^{10}) \to \mathcal{A}_\theta$ by
$f \mapsto \mathcal{Q}(f) \equiv \widehat{f}$ is a $\mathbb{C}$-linear algebra homomorphism
defined by
\begin{equation}\label{q-rule}
    f \cdot  g \mapsto \widehat{f \star g} = \widehat{f} \cdot \widehat{g}
\end{equation}
and
\begin{equation}\label{quantum-prod}
   f \star g \equiv \mathcal{Q}^{-1} \Bigl( \mathcal{Q}(f) \cdot \mathcal{Q}(g) \Bigr)
\end{equation}
for $f, g \in C^\infty(\mathbb{R}^{10})$ and $\widehat{f}, \widehat{g} \in \mathcal{A}_\theta$.
The above star product is given by Eq. \eq{10d-star-prod} \cite{ncft-rev}. The DBI action \eq{ncdbi}
for the NC D9-brane relevant to the NC $U(1)$ gauge theory \eq{10dsym-action} is then obtained by
simply considering a particular NC parameter $\Theta^{MN} = (\zeta^{\mu\nu}, \theta^{ab})$
with $\zeta^{\mu\nu} = 0$. We understand the limit $\zeta^{\mu\nu} \to 0$ as $|\zeta|^2
\equiv G_{\mu\rho} G_{\nu\sigma} \zeta^{\mu\nu} \zeta^{\rho\sigma} = \kappa^2 |\kappa B_{\mu \lambda}
g^{\lambda\rho}|^2 \ll \kappa^2$ where the open string metric in Eq. \eq{metric-coupling} was used.
This means that $g_{\mu\nu} + \kappa B_{\mu\nu} = (\delta^\rho_\mu + \kappa B_{\mu\lambda}
g^{\lambda\rho}) g_{\rho\nu} \approx g_{\mu\nu}$, in other words,
the metric part in the DBI background $g_{\mu\nu} + \kappa B_{\mu\nu}$ is dominant so that
the $B$-field part can be ignored.

Why do we need to take the limit $\zeta^{\mu\nu} \to 0$ instead of simply putting $\zeta^{\mu\nu} = 0$?
Actually the answer is involved with the most beautiful aspect of emergent gravity.
In the emergent gravity picture, any spacetime structure is not assumed {\it a priori} but
defined by the theory itself. In a sonorous phrase, the theory of emergent gravity must be
background independent. Hence it is necessary to define a configuration in the algebra $\mathcal{A}_\theta$,
for instance, like Eq. (\ref{extra-nc2n}), to generate any kind of spacetime structure,
even for flat spacetime. Emergent gravity then says that the flat spacetime is emergent from
the Moyal-Heisenberg algebra (\ref{extra-nc2n}).
In other words, even the flat spacetime must have a dynamical origin \cite{mypaper,hsy-jhep09,hsy-jpcs12},
which is absent in general relativity.
This picture may also be convinced by gazing up at the identity \eq{dbi-idc}.
Note that the dynamical variables on the RHS of Eq. \eq{dbi-idc} are
(emergent) metric fields, $\mathcal{G}_{MN} (Y)$, whereas they on the left-hand side (LHS) are
$U(1)$ gauge fields, $F_{MN} (X)$, in a specific background $(g,B)$.
Therefore the gravitational fields $\mathcal{G}_{MN} (Y)$ are completely determined by dynamical $U(1)$
gauge fields and so the former is emergent from the latter. When $U(1)$ gauge fields are turned off,
the emergent metric reduces to the flat metric, i.e., $\mathcal{G}_{MN} = g_{MN}$.
But the background $B$-field still persists and it can be regarded
as a vacuum gauge field $A^{(0)}_M  = - \frac{1}{2} B_{MN} X^N$.
Then it is natural to think that the flat metric $g_{MN}$ is emergent from the vacuum gauge fields $A^{(0)}_M$.
This remarkable picture can be rigorously confirmed from a background independent formulation,
e.g., matrix models \cite{mypaper,hsy-jhep09,hsy-jpcs12}. In consequence, any spacetime
structure did not exist {\it a priori} but the existence of spacetime requires
a coherent condensate of vacuum gauge fields. Nature allows ``no free lunch."
As a result, the usual commutative spacetime
has to be understood as a {\it commutative} limit of NC spacetime as we advocated above.
Indeed we do not know how to reproduce the NC DBI action \eq{ncdbi} via the identity \eq{dbi-idc}
starting with the $U(1)$ field strength \eq{matrix-10f}.\footnote{Note that
the Darboux theorem \eq{darboux-local} can be applied only to a symplectic form, i.e.,
a nondegenerate and closed 2-form. But the dynamical 2-form $F$ does not belong to this category
because it usually vanishes at an asymptotic infinity.}

Note that the coordinate transformation \eq{darboux-local} to a Darboux frame is defined only locally
and symplectic or NC gauge fields have been introduced to compensate local deformations of
an underlying symplectic structure by $U(1)$ gauge fields, i.e., the Darboux coordinates in $\phi:Y
\mapsto X=X(Y) \in \mathrm{Diff}(\mathbb{R}^{10})$ obey the relation $\phi^* (B+F) = B$.
The identity (\ref{rexp-dbi}) also manifests this local nature of NC gauge fields because
they manifest themselves only in a locally inertial frame (in free fall) with the
local metric (\ref{de-metric}) \cite{mypaper}. If the gravitational metric in Eq. (\ref{rexp-dbi})
were represented by a global form, e.g.,
\begin{equation}\label{global-metric}
    \mathcal{G}_{MN} = g_{AB} E^A_M E^B_N, \qquad A, B = 0, 1, \cdots, 9
\end{equation}
where $E^A = E^A_M dx^M$ are elements of a global coframe on an emergent 10-dimensional
manifold $\mathcal{M}$, it would be difficult to find an imprint of symplectic or NC gauge
fields in the expression \eq{global-metric}.

Recall that the basic program of differential geometry is that all the world can be reconstructed
from the infinitely small. For example, manifolds are obtained by gluing open subsets of Euclidean space.
So the differential forms and vector fields on a manifold are defined locally and then glued together
to yield a global object. The gluing is possible because these objects are independent of the choice
of local coordinates. In reality this kind of globalization of a (spacetime) geometry by
gluing local data might be enforced because global comparison devices are not available
owing to the restriction of the finite propagation speed.
Indeed the global metric \eq{global-metric} can be constructed in a similar way.
First note that the D9-brane described by the LHS of Eq. \eq{dbi-idc} supports
a line bundle $L \to \mathbb{R}^{10}$ over a symplectic manifold $(\mathbb{R}^{10}, B)$.
Introduce an open covering $\{U_i: i \in I \}$ of $\mathbb{R}^{10}$, i.e.,
$\mathbb{R}^{10} = \bigcup_{i \in I} U_i$ and let $A^{(i)}$ be a connection of the line
bundle $L \to U_i$ on an open neighborhood $U_i$.
Consider all compatible coordinate systems $\{ (U_i, \varphi_i): i \in I \}$
as a family of local Darboux charts where $\varphi_i: U_i \to \mathbb{R}^{10}$ are
Darboux coordinates on $U_i$. Then we have the collection of
local data $\bigoplus_{i \in I}(A^{(i)}, Y_{(i)})$ on the D9-brane
where $Y_{(i)} = \varphi_i (U_i)$ are Darboux coordinates on $U_i$ obeying Eq. \eq{darboux-local},
i.e., $\varphi_i^* (B + F^{(i)}) =B$ where $F^{(i)} = d A^{(i)}$.
On an intersection $U_i \cap U_j$, local data $(A^{(i)}, Y_{(i)})$ and $(A^{(j)}, Y_{(j)})$
on Darboux charts $U_i$ and $U_j$, respectively, are glued together by \cite{jsw-ncl,buba-me}
\begin{eqnarray} \label{glue-g}
&& A^{(j)} = A^{(i)} + d\lambda^{(ji)}, \\
\label{glue-d}
&& Y_{(j)} = \varphi_{(ji)} (Y_{(i)}),
\end{eqnarray}
where $\varphi_{(ji)}$ is a symplectomorphism on $U_i \cap U_j$ generated by a Hamiltonian
vector field $X_{\lambda^{(ji)}}$ obeying $\iota_{X_{\lambda^{(ji)}}} B  + d \lambda^{(ji)} = 0$.
Note that the symplectomorphism is a canonical transformation preserving
the Poisson structure $\Theta = B^{-1}$ and can be identified with a NC $U(1)$
gauge transformation upon quantization \cite{hsy-ijmp09,ncft-rev}.
Since the local metric (\ref{de-metric}) is the incarnation of
symplectic gauge fields in a Darboux frame, the gluing of local Darboux charts can be translated into
that of emergent metrics in locally inertial frames from the viewpoint of the RHS
of Eq. \eq{dbi-idc}. This kind of gluing should be well-defined because every manifold can be constructed
by gluing open subsets of Euclidean space together and both sides of Eq. \eq{dbi-idc} are coordinate
independent and so local Darboux charts can be consistently glued altogether. See Ref. \cite{lry3}
to illuminate how a nontrivial topology of an emergent manifold can be implemented by gluing local
data $\bigcup_{i \in I}(A^{(i)}, Y_{(i)})$.

It is in order to ponder on the results obtained. We showed in section 2 that the 4-dimensional
$\mathcal{N}=4$ super Yang-Mills theory on the Coulomb branch \eq{n=4vacuum} is equivalent to
the 10-dimensional $\mathcal{N}=1$ supersymmetric NC $U(1)$ gauge theory. And we considered
the resulting 10-dimensional NC $U(1)$ gauge theory as a low-energy effective theory
of supersymmetric NC D9-brane. Finally we got the important identity (\ref{rexp-dbi})
that the dynamics of NC $U(1)$ gauge fields after ignoring fermion fields is completely encoded
into a 10-dimensional emergent geometry described by the metric \eq{global-metric}.
According to the AdS/CFT correspondence, it is natural to expect that the metric \eq{global-metric}
must describe a 10-dimensional emergent geometry dual to the 4-dimensional $\mathcal{N}=4$ super
Yang-Mills theory. An immediate question to arise is how to realize the $AdS_5 \times \mathbb{S}^5$
vacuum geometry in our context.

Since there is no reason to further reside in Euclidean space, let us go back to the Lorentzian
spacetime with the NC parameter $\Theta^{MN} = (\zeta^{\mu\nu} = 0, \theta^{ab} \neq 0)$ by Wick rotation.
In order to pose the above question, let us consider a more general vacuum geometry which
is conformally flat. That is, we are interested in a background geometry with the metric given by
\begin{equation}\label{10vacgeo}
    ds^2 = \lambda^2 (\eta_{\mu\nu} dx^\mu dx^\nu + dy^a dy^a).
\end{equation}
There are two interesting cases which are conformally flat \cite{mypaper}:
\begin{eqnarray} \label{vacgeo1}
&& \lambda^2=1 \qquad \Rightarrow \quad \mathcal{M} = \mathbb{R}^{9,1}, \\
\label{vacgeo2}
&& \lambda^2=\frac{R^2}{\rho^2} \quad \; \Rightarrow \quad \mathcal{M} = AdS_5 \times \mathbb{S}^5,
\end{eqnarray}
where $\rho^2 = \sum_{a=1}^6 y^a y^a$ and $R = \bigl (4 \pi g_s (\alpha')^2 N \bigr)^{1/4}$ is
the radius of $AdS_5$ and $\mathbb{S}^5$ spaces.
We already speculated before that the flat Minkowski
spacetime (\ref{vacgeo1}) arises from a uniform condensate of vacuum gauge fields
$A^{(0)}_M  = - \frac{1}{2} B_{MN} X^N$. This can be confirmed by looking at the vacuum
configuration \eq{n=4vacuum}. Note that, from the 4-dimensional gauge theory point of view,
the vacuum configuration \eq{n=4vacuum} simply represents a particular configuration of large $N$ matrices
and it is connoted as an extra 6-dimensional ``emergent" space only in 10-dimensional description.
Its tangible existence must be addressed from the RHS of Eq. \eq{dbi-idc}.
(See section 1 in Ref. \cite{mypaper} for the rationale underlying this reasoning.)
Then it is easy to prove that the emergent metric \eq{de-metric} for the vacuum
configuration \eq{n=4vacuum} is precisely the flat Minkowski spacetime (\ref{vacgeo1}).
Note that a Darboux chart $(U, \varphi)$ in this case can be extended to entire spacetime and
so it is not necessary to consider the globalization prescribed before.

Now a perplexing problem is to understand what is the gauge field configuration to realize
the vacuum geometry \eq{vacgeo2}. In order to figure out the problem,
it is necessary to find a stable configuration of NC or large $N$ gauge fields and
so certainly a supersymmetric or BPS state. And this configuration must be consistent with the isometry
of the vacuum geometry (\ref{10vacgeo}), in particular, preserving $SO(6)_R$ Lorentz symmetry
as if a hydrogen atom preserves $SO(3)$ symmetry. It was conjectured in \cite{mypaper} that
the $AdS_5 \times \mathbb{S}^5$ geometry arises from the stack of NC Hermitian $U(1)$ instantons
at origin in the internal space $\mathbb{R}^6$ like a nucleus containing a lot of nucleons.
The NC Hermitian $U(1)$ instanton obeys the Hermitian Yang-Mills equations \cite{non-inst} given by
\begin{equation}\label{hym-eq}
    \widehat{F}_{ab} = - \frac{1}{4} \varepsilon_{abcdef} \widehat{F}_{cd} I_{ef},
\end{equation}
where $I = \mathbf{I}_3 \otimes i \sigma^2$ is a $6 \times 6$ matrix of the complex structure
of $\mathbb{R}^6$ and the field strength is defined by Eq. \eq{10d-fs}.
Note that the 6-dimensional NC $U(1)$ gauge fields $\widehat{A}_a$ in Eq. \eq{hym-eq} are originally
adjoint scalar fields $\Phi_a = p_a + \widehat{A}_a$ in 4-dimensional $\mathcal{N} = 4$
super Yang-Mills theory. See Eq. \eq{n=4bfluct}. If true, the vacuum geometry \eq{vacgeo2}
will be emergent from the stack of infinitely many NC $U(1)$ instantons obeying Eq. \eq{hym-eq}
according to the identity \eq{rexp-dbi}.\footnote{Given the metric \eq{10vacgeo} of
$AdS_5 \times \mathbb{S}^5$ geometry on the LHS of Eq. \eq{rexp-dbi},
we may simply assume that we have solved Eq. \eq{rexp-dbi} to find some configuration of $U(1)$ gauge fields
which gives rise to the $AdS_5 \times \mathbb{S}^5$ geometry. In appendix A, we will solve Eq. \eq{rexp-dbi}
to illustrate how $2n$-dimensional Calabi-Yau manifolds arise from $2n$-dimensional symplectic $U(1)$
gauge fields. But it should be remarked that the underlying argument can proceed with impunity
whatever our conjecture is true or not.}
Since we are interested in the approximation of
slowly varying fields, $\sqrt{\theta} |\frac{\widehat{D} \widehat{F}}{\widehat{F}}| \ll 1$,
ignoring the derivatives of field strengths, the $U(1)$ field strength in Eq. \eq{hym-eq}
can be replaced by Eq. \eq{symp-f} in this limit and so we can use
the SW maps \eq{sw-mapf} and \eq{sw-mapv}. Thus, if we include NC corrections containing
higher-order derivatives of field strengths, the LHS of Eq. \eq{rexp-dbi} will
receive derivative corrections introducing a higher-order gravity
in the emergent geometry \cite{hsy-ijmp09}.

In conclusion, the AdS/CFT correspondence is a particular example of emergent gravity
from NC $U(1)$ gauge fields. And the duality between large $N$ gauge fields and
a higher-dimensional gravity is simply a consequence of the novel equivalence principle
stating that the electromagnetic force can always be eliminated by a local coordinate transformation
as far as spacetime admits a symplectic structure, in other words, a microscopic spacetime
becomes NC \cite{mypaper,hsy-jhep09}.

\section{HEA from NC $U(1)$ gauge fields}

Now we are ready to derive the HEA of four-dimensional $\mathcal{N}=4$ superconformal field theory
on the Coulomb branch. According to the conjecture \cite{schwarz}, the HEA should be
a $U(1)$ gauge theory in the $AdS_5 \times \mathbb{S}^5$ geometry with $N$ units of flux
threading $\mathbb{S}^5$. However the original conjecture did not allude any clue
why the HEA on the Coulomb branch must be described by the $U(1)$ gauge theory although
the probe-brane approximantion requires a large $N$ limit. For the discussion of this problem,
see, in particular, section 5 in Ref. \cite{schwarz}. As we emphasized in footnote \ref{n=1},
our approach based on the NC field theory representation of AdS/CFT correspondence
will clarify why $N=1$ is the relevant choice for the HEA.

We argued before that the $AdS_5 \times \mathbb{S}^5$ geometry is emergent from the stack of
infinitely many NC Hermitian $U(1)$ instantons near origin in $\mathbb{R}^6$.
Thus suppose that the vacuum configuration for the background geometry \eq{vacgeo2} is given by
\begin{equation}\label{inst-vacuum}
    \langle \Phi_a \rangle_{\mathrm{vac}} = p_a + \widehat{A}_a, \quad
    \langle A_\mu \rangle_{\mathrm{vac}} = 0, \quad \langle \lambda^i \rangle_{\mathrm{vac}} = 0,
\end{equation}
where $\widehat{A}_a$ is a solution of Eq. \eq{hym-eq} describing $N$ NC Hermitian $U(1)$ instantons
in 6 dimensions. We introduce fluctuations around the vacuum \eq{inst-vacuum} and represent them as
\begin{eqnarray}\label{conn-fluc4}
    && \widehat{D}_\mu = \partial_\mu - i \widehat{a}_\mu (x,y), \\
    \label{conn-fluc6}
    && \widehat{D}_a = - i \bigl( p_a  + \widehat{A}_a (y)  + \widehat{a}_a  (x,y) \bigr)
    \equiv \widehat{\nabla}_a (y) - i  \widehat{a}_a  (x,y),
\end{eqnarray}
whose field strengths are given by
\begin{eqnarray}\label{f-fluc44}
\widehat{\mathcal{F}}_{\mu\nu} &=& \partial_\mu \widehat{a}_\nu - \partial_\nu \widehat{a}_\mu - i
    [\widehat{a}_\mu, \widehat{a}_\nu]_\star \equiv \widehat{f}_{\mu\nu}, \\
    \label{conn-fluc46}
\widehat{\mathcal{F}}_{\mu a} &=& \widehat{D}_\mu \widehat{a}_a - \widehat{\nabla}_a \widehat{a}_\mu \equiv \widehat{f}_{\mu a}, \\
    \label{conn-fluc66}
\widehat{\mathcal{F}}_{ab} &=& - B_{ab} + \widehat{F}_{ab}
    + \widehat{\nabla}_a \widehat{a}_b - \widehat{\nabla}_b \widehat{a}_a
    - i [\widehat{a}_a, \widehat{a}_b]_\star, \nonumber\\
    &\equiv& - B_{ab} + \widehat{F}_{ab} + \widehat{f}_{ab}
\end{eqnarray}
where $\widehat{F}_{ab} (y) - B_{ab} = i [\widehat{\nabla}_a, \widehat{\nabla}_b]_\star (y)$.
We will include fermions later.
Note that we assumed that the instanton connection $\widehat{\nabla}_a (y)$ depends only
on NC coordinates in extra dimensions. Hence the solution has a translational invariance
along $\mathbb{R}^{3,1}$ which means that the solution describes extended objects
along $\mathbb{R}^{3,1}$. They were conjecturally identified with $N$ D3-branes in \cite{mypaper}.
Since the SW relation between commutative and NC gauge theories is true for general gauge fields,
we can apply to the gauge fields in Eqs. \eq{conn-fluc4} and \eq{conn-fluc6} the SW maps
\begin{eqnarray}\label{sw-instf}
    && \widehat{\mathcal{F}}_{MN} (Y) = \left( \frac{1}{1 + \mathfrak{F}\Theta} \mathfrak{F}
    \right)_{MN} (X), \\
    \label{sw-instv}
    &&  d^{10} Y =  d^{10} X \sqrt{\det(1 + \mathfrak{F}\Theta)},
\end{eqnarray}
where $\mathfrak{F} \equiv B + F + f$ is the total $U(1)$ field strength including the background
instanton part $F_{ab}$ and the fluctuation part $f_{MN} = \partial_M a_N - \partial_N a_M$.
The result will be given by the following equivalence
\begin{equation} \label{dbi-fluc}
\frac{1}{g_s} \int d^{10} X \sqrt{ - \det \bigl( g + \kappa \mathfrak{F} \bigr)}
= \frac{1}{G_s} \int d^{10} Y \sqrt{- \det \bigl(G  + \kappa (\widehat{\mathcal{F}} + \Phi ) \bigr)}.
\end{equation}
But we can also apply the Darboux transformation \eq{darboux-local} to the field strength $\mathfrak{F}$
such that the Darboux coordinates $Z^M$ eliminate only the instanton gauge fields $F_{ab}$.
Then we will get the following identity
\begin{equation}\label{tri-dbi}
    g_{MN} + \kappa \mathfrak{F}_{MN} = \bigl(\mathcal{G}_{PQ} + \kappa (B + \widetilde{f})_{PQ} \bigr)
    \frac{\partial Z^P}{\partial X^M} \frac{\partial Z^Q}{\partial X^N}
\end{equation}
where
\begin{equation}\label{tr-abelif}
\mathcal{G}_{MN} = g_{PQ} \frac{\partial X^P}{\partial Z^M}
 \frac{\partial X^Q}{\partial Z^N}, \qquad
 \widetilde{f}_{MN} = f_{PQ} \frac{\partial X^P}{\partial Z^M} \frac{\partial X^Q}{\partial Z^N}
    = \frac{\partial \widetilde{a}_N}{\partial Z^M}  - \frac{\partial \widetilde{a}_M}{\partial Z^N}
\end{equation}
with $\widetilde{a}_M = \frac{\partial X^P}{\partial Z^M} a_P$. This leads to an enticing result
\begin{eqnarray} \label{dbi-idif}
\frac{1}{g_s} \int d^{10} X \sqrt{-\det \bigl( g + \kappa \mathfrak{F} \bigr)}
&=& \frac{1}{g_s} \int d^{10} Z \sqrt{-\det \bigl(\mathcal{G}  + \kappa (B + \widetilde{f}) \bigr)} \\
 \label{dbi-idin}
&=& \frac{1}{G_s} \int d^{10} Y \sqrt{-\det \bigl(G  + \kappa (\widehat{\mathcal{F}} + \Phi ) \bigr)}.
\end{eqnarray}

We can check the consistency of the above identities by showing that Eq. \eq{dbi-idin} can be derived
from the RHS of Eq. \eq{dbi-idif}.
Consider a Darboux transformation $\phi_1: Y^M \mapsto Z^M = Y^M + \Theta^{MN} \widehat{a}_N (Y)$
satisfying $\phi_1^* (B + \widetilde{f}) = B$. Then it leads to the identity
\begin{equation}\label{inter-dar}
\mathcal{G}_{MN} + \kappa (B + \widetilde{f})_{MN} =  \bigl( \mathfrak{G}_{PQ} + \kappa B_{PQ} \bigr)
\frac{\partial Y^P}{\partial Z^M} \frac{\partial Y^Q}{\partial Z^N}
\end{equation}
where
\begin{equation}\label{inter-metric}
\mathfrak{G}_{MN} = \mathcal{G}_{PQ} \frac{\partial Z^P}{\partial Y^M}
 \frac{\partial Z^Q}{\partial Y^N} = g_{PQ} \frac{\partial X^P}{\partial Y^M}
 \frac{\partial X^Q}{\partial Y^N}.
\end{equation}
The previous Darboux transformation \eq{tri-dbi} satisfies $\phi_2^* (B + F) = B$ where
$\phi_2: Z^M \mapsto X^M = Z^M + \Theta^{MN} \widehat{A}_N (Z)$ which, in Eq. \eq{inter-metric},
has been combined with $\phi_1$, i.e.,
\begin{equation}\label{comb-darb}
    \phi_2 \circ \phi_1: Y^M \mapsto X^M = Y^M + \Theta^{MN} (\widehat{A}_N + \widehat{a}_N ) (Y).
\end{equation}
Note that we can put $\widehat{A}_\mu = 0$ by our assumption.
Using the identity \eq{inter-dar}, we can derive the following equivalence between DBI actions:
\begin{equation} \label{dbi-ccc}
\frac{1}{g_s} \int d^{10} Z \sqrt{- \det \bigl(\mathcal{G}  + \kappa (B + \widetilde{f}) \bigr)}
= \frac{1}{g_s} \int d^{10} Y \sqrt{- \det \bigl( \mathfrak{G} + \kappa B \bigr)}.
\end{equation}
By applying the same method as Eq. \eq{rexp-dbi} and using the coordinates \eq{comb-darb},
it is straightforward to derive Eq. \eq{dbi-idin} from the RHS of Eq. \eq{dbi-ccc}.

The conformally flat metric \eq{10vacgeo} takes the form
\begin{equation}\label{asd2s5}
    ds^2 = R^2 \Bigl(\frac{dx \cdot dx + d\rho^2}{\rho^2} + d \Omega_5^2 \Bigr)
\end{equation}
where $dx \cdot dx = \eta_{\mu\nu} dx^\mu dx^\nu$. This form of the metric can be transformed into
the metric form used in \cite{schwarz} by a simple inversion $\rho = 1/ v$:
\begin{equation}\label{metric-sch}
    ds^2 = R^2 \Bigl(v^2 dx \cdot dx + v^{-2} dv^2 + d \Omega_5^2 \Bigr) =
    R^2 \Bigl(v^2 dx \cdot dx + v^{-2} dv \cdot dv \Bigr)
\end{equation}
where $dv \cdot dv = dv^a dv^a$.
Note that the four-dimensional supersymmetric gauge theory is defined on the boundary of $AdS_5$ space
where $v \to \infty$ in the metric (\ref{metric-sch}) and so the five-sphere $\mathbb{S}^5$
shrinks to a point near the conformal boundary of the AdS space.
Then the $SO(6)$ isometry of $\mathbb{S}^5$ is realized as a global symmetry in the gauge theory
and the (angular) momenta dual to five-sphere coordinates are given by generators of
the $SO(6)$ R-symmetry. Since we are interested in the HEA of the boundary theory where
the $\mathbb{S}^5$ shrinks to a point, we can thus consider a low energy limit by ignoring any $y$-dependence
for fluctuations, but leaving the background intact. Then the fluctuating $U(1)$ field strengths
on the LHS of Eq. \eq{dbi-ccc} reduce to
\begin{equation}\label{lowel-u1}
\begin{array}{l}
  \widetilde{f}_{\mu\nu} (x,y) \to \partial_\mu \widetilde{a}_\nu (x) - \partial_\nu \widetilde{a}_\mu (x)
  \equiv f_{\mu\nu}(x), \\
  \widetilde{f}_{\mu a} (x,y) \to \partial_\mu \widetilde{a}_a (x) \equiv \partial_\mu \varphi_a (x), \\
  \widetilde{f}_{ab} (x,y) \to 0.
\end{array}
\end{equation}
Since we assumed that the low energy theory does not depend on
the coordinates $y^a$ of extra dimensions, we will try to reduce the 10-dimensional theory to a
4-dimensional effective field theory. For this purpose, first let us consider the block matrix
\begin{equation}\label{bdi-10metric}
\mathcal{G}_{MN} + \kappa \bigl(B + \widetilde{f} \bigr)_{MN} = \left(
  \begin{array}{cc}
    \lambda^2 \eta_{\mu\nu} + \kappa f_{\mu\nu} & \kappa \partial_\mu \varphi_a \\
    - \kappa \partial_\mu \varphi_a & \lambda^2 \delta_{ab} + \kappa B_{ab} \\
  \end{array}
\right),
\end{equation}
where we put $B_{\mu\nu} = 0$ according to the reasoning explained in section 4.
Even we may take the approximation $\lambda^2 \delta_{ab} + \kappa B_{ab} \approx \lambda^2 \delta_{ab}$
because $\lambda^2 = R^2 v^2 \to \infty$ and the low energy limit applied to Eq. \eq{lowel-u1}
is basically equivalent to $\theta^{ab} \to 0$ and so the metric part is dominant similarly to the reasoning
below Eq. \eq{quantum-prod}. Considering the fact that NC corrections in NC gauge theory correspond to
$1/N$ expansions in large $N$ gauge theory \cite{hsy-ijmp09},
the approximation considered can be interpreted as the planar limit in AdS/CFT correspondence.
Using the determinant formula for a block matrix
\begin{equation}\label{det-block}
    \det \left(
           \begin{array}{cc}
             A & B \\
             C & D \\
           \end{array}
         \right) = \det D \; \det (A-B D^{-1} C),
\end{equation}
we get the following relation
\begin{eqnarray} \label{prod-det}
\sqrt{- \det \bigl(\mathcal{G}  + \kappa (B + \widetilde{f}) \bigr)} &=&
\sqrt{\det (\lambda^2 + \kappa B )} \sqrt{- \det \Bigl( \lambda^2 \eta_{\mu\nu}
+ \kappa f_{\mu\nu}
+ \kappa^2 \partial_\mu \varphi_a \Bigl(\frac{1}{\lambda^2 + \kappa B} \Bigr)^{ab}
\partial_\nu \varphi_b \Bigr) } \nonumber \\
  &\approx& \lambda^6 \sqrt{- \det \Bigl( \lambda^2 \eta_{\mu\nu} + \kappa^2 \lambda^{-2} \partial_\mu \varphi
  \cdot \partial_\nu \varphi + \kappa f_{\mu\nu} \Bigr) }.
\end{eqnarray}

Suppose that a D3-brane is embedded in 10-dimensional target spacetime $\mathcal{M}$ with local coordinates
$X^M = (x^\mu, \phi^a)$ whose metric is given by $\mathcal{G}_{MN} (X)$. To be specific, we consider
$\mathcal{M} = AdS_5 \times \mathbb{S}^5$ and choose a static gauge for the embedding functions, i.e.,
$X^M (\sigma) = \bigl( x^\mu(\sigma), \phi^a(\sigma) \bigr) = \bigl( \delta^\mu_\alpha \sigma^\alpha,
v^a + \frac{\kappa}{R^2} \varphi^a(x) \big)$ where $v^a \equiv \langle \phi^a \rangle_{\mathrm{vac}}$
are vevs of worldvolume scalar fields. The fact that the worldvolume scalar fields $\phi^a$ are
originated from NC $U(1)$ gauge fields in Eq. \eq{conn-fluc6} implies that the vevs
$v^a = \langle \phi^a \rangle_{\mathrm{vac}}$ can be identified with the Coulomb branch
parameters $p_a$ in Eq. \eq{n=4vacuum}.
Then we see that the symmetric part in Eq. \eq{prod-det} is precisely the induced worldvolume
metric \eq{ind-metric}, i.e.,
\begin{equation}\label{d3-indmetric}
    h_{\mu\nu} = \mathcal{G}_{MN} \partial_\mu X^M \partial_\nu X^N
= R^2 \bigl( v^2 \eta_{\mu\nu} + v^{-2} \partial_\mu \phi \cdot \partial_\nu \phi  \bigr)
\end{equation}
where $\lambda^2 = R^2 v \cdot v = R^2 /\rho^2$.
Therefore, in the approximation considered above, we get the identity
\begin{equation}\label{hea-id}
 \sqrt{- \det_{10} \bigl(\mathcal{G}  + \kappa (B + \widetilde{f}) \bigr)} =
 \lambda^6 \sqrt{- \det_4 (h  + \kappa f)}
\end{equation}
where the subscript in the determinant indicates the size of matrix.
Using the identity \eq{hea-id}, we can reduce the 10-dimensional DBI action in $AdS_5 \times \mathbb{S}^5$
geometry to a 4-dimensional DBI action given by
\begin{equation}\label{hea-red}
- T_{D9} \int d^{10} Z  \sqrt{- \det_{10} \bigl(\mathcal{G}  + \kappa (B + \widetilde{f}) \bigr)} =
 \Bigl( \frac{g_s N}{4\pi}\Bigr)^{\frac{3}{2}} L(\epsilon, R)
 \left[ - T_{D3} \int_W d^4 x \sqrt{- \det_4 (h  + \kappa f)} \right]
\end{equation}
where $\Bigl( \frac{g_s N}{4\pi}\Bigr)^{\frac{3}{2}} = \frac{T_{D9} R^6}{T_{D3}} \int_{\mathbb{S}^5} \mathrm{vol}(\mathbb{S}^5)$ and
\begin{equation}\label{reg-radius}
L(\epsilon, R) \equiv   \int^R_\epsilon \frac{dv}{v} = \ln \frac{R}{\epsilon}
\end{equation}
is a regularized integral along the $AdS$ radius. We identify the DBI action in the bracket
in Eq. \eq{hea-red} with the worldvolume action of a probe D3-brane in $AdS_5 \times \mathbb{S}^5$ geometry.
John H. Schwarz speculated in \cite{schwarz} that the probe D3-brane action can be interpreted as the HEA
of 4-dimensional $\mathcal{N}=4$ superconformal field theory on the Coulomb branch.
We want to emphasize we directly derived the HEA from the 4-dimensional $\mathcal{N}=4$ superconformal
field theory on the Coulomb branch although we have not incorporated fermions yet.
One caveat is that our HEA is slightly different from Eq. (12) in Ref. \cite{schwarz} where our $v^2$
was replaced by $\phi^2$. But one needs to recall that $v^2$ is coming from the background geometry and
the probe brane approximation involves neglecting the backreaction of the brane on the geometry and
other background fields (which requires that $N$ is large).
In this description, the $AdS_5 \times S^5$ geometry is regarded as a background and
so it remains to be fixed against the fluctuations of worldvolume fields.
Thus the $\phi^2$ in the denominator in Eq. (12) of Ref. \cite{schwarz} can be replaced by $v^2$
in the probe brane approximation.

A demanding task is to understand how to derive the coupling \eq{wz-action} of background
RR gauge fields from the 4-dimensional $\mathcal{N}=4$ superconformal field theory.
Actually this issue is closely related to our previous conjecture for a possible realization of
D3-branes in terms of NC Hermitian $U(1)$ instantons. Hence we will only draw a plausible picture
based on this conjecture. If the conjecture is true, $N$ D3-branes correspond
to a stack of $N$ NC Hermitian $U(1)$ instantons at origin of $\mathbb{R}^6$. Then,
this instanton configuration generates a topological invariant given by (up to normalization)
\begin{equation}\label{top-number}
    I \sim \int_{\mathbb{R}^6} \widehat{F} \wedge \widehat{F} \wedge \Omega =
    \int_{\mathbb{S}^5} \Bigl(\widehat{A} \wedge \widehat{F}
    - \frac{1}{3} \widehat{A} \wedge \widehat{A} \wedge  \widehat{A} \Bigr) \wedge \Omega
\end{equation}
where $\Omega$ is a K\"ahler form on $\mathbb{R}^6$. The topological invariant $I$ refers to
the instanton number $N$ and so we identify $I = 2 \pi N$. Since the ``instanton flux" is
threading $\mathbb{S}^5 = \partial \mathbb{R}^6$ and the instanton flux emanating from the origin
is regarded as a background field, we make a simple identification for the five-form
in Eq. \eq{top-number}:
\begin{eqnarray}\label{5-form}
    \mu_3 F_5 &:=& \frac{1}{g_{YM}^2} \Bigl(\widehat{A} \wedge \widehat{F}  - \frac{1}{3} \widehat{A} \wedge
    \widehat{A} \wedge  \widehat{A} \Bigr) \wedge \Omega \nonumber \\
    &=& \mu_3 k_3 \mathrm{vol} (\mathbb{S}^5)
\end{eqnarray}
where $\mu_3$ is the basic unit of D3-brane charge and $k_3$ is a coefficient depending
on the normalization convention. In the AdS/CFT correspondence, $F_5$ is the self-dual RR five-form of
$N$ D3-branes given by
\begin{equation}\label{rr-5form}
    F_5 = k_3 \bigl( \mathrm{vol} (AdS_5) + \mathrm{vol} (\mathbb{S}^5) \bigr) = dC_4.
\end{equation}
Although we do not pin down the origin of the self-duality, the self-duality is necessary
for the conjecture to be true because it implies that the topological charge of NC $U(1)$ instantons
can be interpreted as the RR-charge of D3-branes, i.e.,
\begin{equation}\label{d3-rrc}
    \mu_3 \int_{\mathbb{S}^5} F_5 = \mu_3 \int_{AdS_5} dC_4 = \mu_3 \int_W C_4
\end{equation}
where $W = \partial (AdS_5)$. Besides the background instanton gauge fields,
there exist worldvolume $U(1)$ gauge fields and they can induce a well-known topological instanton coupling
given by
\begin{equation}\label{w-instanton}
    \frac{\chi}{8\pi} \int_W f \wedge f.
\end{equation}
 Combining these two couplings leads to a moderate (if any) suggestion
for the Wess-Zumino coupling in Eq. \eq{wz-action} given by \cite{schwarz}
\begin{equation}\label{scs2}
    S_2 = \mu_3 \int_W C_4 +  \frac{\chi}{8\pi} \int_W f \wedge f.
\end{equation}

Now we will include the Majorana-Weyl fermion $\widehat{\Psi}(Y)$ in the HEA.
This means that we are considering a supersymmetric D9-brane which respects the
local $\kappa$-symmetry \cite{sdbi1,sdbi12,sdbi3,sdbi4,sdbi5,sdbi6}.
Thus we use the $\kappa$-symmetry to eliminate half of $(\psi_1, \psi_2)$ coordinates
where $\psi_{1,2}$ are two Majorana-Weyl spinors of the same chirality.
We adopt the gauge choice, $\psi_1 = 0$, used in Ref. \cite{sdbi1,sdbi12} and rename $\psi_2 := \psi$.
It was shown in \cite{sdbi1,sdbi12} that in this gauge the supersymmetric extension of 10-dimensional DBI action
has a surprisingly simple form. The supersymmetric case also respects the identity \eq{dbi-idif}
with the following replacement
\begin{eqnarray}\label{susy-rep1}
    && \mathfrak{F}_{MN} \to  \mathfrak{F}_{MN} + i \overline{\psi} \Gamma_M \partial_N \psi -
    \frac{\kappa}{4} \overline{\psi} \Gamma^P \partial_M \psi \overline{\psi} \Gamma_P \partial_N \psi  \equiv
    \mathfrak{F}_{MN} + \Upsilon_{MN}, \\
    \label{susy-rep2}
    && \widetilde{f}_{MN} \to  \widetilde{f}_{MN} + i \overline{\psi} \widetilde{\Gamma}_M
    \widetilde{\partial}_N \psi - \frac{\kappa}{4} \overline{\psi} \widetilde{\Gamma}^P \widetilde{\partial}_M
    \psi \overline{\psi} \widetilde{\Gamma}_P \widetilde{\partial}_N \psi \equiv \widetilde{f}_{MN} + \xi_{MN},
\end{eqnarray}
where $\widetilde{\Gamma}_M = \Gamma_P \frac{\partial X^P}{\partial Z^M}$ and $\widetilde{\partial}_M = \frac{\partial }{\partial Z^M}$. Again we can apply the Darboux transformation $\phi_1: Y^M \mapsto Z^M =
\Theta^{MN} \bigl( B_{NP} Y^P +  \widehat{a}_N (Y) \bigr)$ satisfying $\phi_1^* (B + \widetilde{f}) = B$.
Then it leads to the following identity
\begin{equation}\label{finter-dar}
\mathcal{G}_{MN} + \kappa (B + \widetilde{f} + \xi)_{MN} =  \bigl( \mathfrak{G}_{PQ}
+ \kappa (B + \widetilde{\xi})_{PQ} \bigr)
\frac{\partial Y^P}{\partial Z^M} \frac{\partial Y^Q}{\partial Z^N}
\end{equation}
where
\begin{equation}\label{finter-metric}
\widetilde{\xi}_{MN} = \xi_{PQ} \frac{\partial Z^P}{\partial Y^M}
 \frac{\partial Z^Q}{\partial Y^N} = \Upsilon_{PQ} \frac{\partial X^P}{\partial Y^M}
 \frac{\partial X^Q}{\partial Y^N}.
\end{equation}
The above identity \eq{finter-dar} leads to the following equivalence between DBI actions:
\begin{equation} \label{dbi-fcc}
\frac{1}{g_s} \int d^{10} Z \sqrt{- \det \bigl(\mathcal{G}  + \kappa (B + \widetilde{f} + \xi) \bigr)}
= \frac{1}{g_s} \int d^{10} Y \sqrt{- \det \bigl( \mathfrak{G} + \kappa (B + \widetilde{\xi}) \bigr)}.
\end{equation}

Let us expand the RHS of Eq. \eq{dbi-fcc} around the background $B$-field as the bosonic case \eq{exp-dbi}:
\begin{equation}\label{exp-fdbi}
 \sqrt{- \det \bigl( \mathfrak{G} + \kappa (B + \widetilde{\xi}) \bigr)}
 = \sqrt{- \det(\kappa B)} \sqrt{\det \Bigl(1 + \frac{M}{\kappa} \Bigr)}
\end{equation}
where
\begin{equation}\label{susy-m}
    {M_N}^Q = \bigl( \mathfrak{G} + \kappa \widetilde{\xi} \bigr)_{NP} \Theta^{PQ}
    = \bigl( g + \kappa \Upsilon \bigr)_{RS} \frac{\partial X^R}{\partial Y^N}
    \frac{\partial X^S}{\partial Y^P} \Theta^{PQ}.
\end{equation}
Note that $\mathrm{Tr} M \neq 0$ unlike the bosonic case.
Using the formula, $\det (1+A) = \exp {\sum_{k=1}^\infty \frac{(-)^{k+1}}{k} \mathrm{Tr} A^k}$,
it is not difficult to show that
\begin{equation}\label{susy-detid}
    \det \Bigl(1 + \frac{M}{\kappa} \Bigr) = \det \Bigl(1 + \frac{1}{\kappa}
    \bigl( g + \kappa \Upsilon \bigr) \mathfrak{P} \Bigr)
\end{equation}
where
\begin{equation}\label{gp-matrix}
 \bigl( \Upsilon \mathfrak{P} \bigr)_M^{~~N} = - i \bigl( \delta^P_M + \frac{i \kappa}{4}
 \overline{\psi} \Gamma^P \partial_M \psi \bigr) \overline{\psi} \Gamma_P \{ X^N, \psi \}_\Theta.
\end{equation}
In terms of the matrix notation, the matrix on the RHS of Eq. \eq{susy-detid} can be read as
\begin{eqnarray} \label{eval-matrix}
   1 + \frac{1}{\kappa} \bigl( g + \kappa \Upsilon \bigr) \mathfrak{P}
   &=& B \bigl(1 + \kappa G^{-1} ( \widehat{\mathcal{F}} - B )
   + \Theta \Upsilon \mathfrak{P} B \bigr) \Theta \nonumber \\
  &=& B  G^{-1} \bigl(G + \kappa ( \widehat{\mathcal{F}} - B )
  + G \Theta \Upsilon \mathfrak{P} B \bigr) \Theta
\end{eqnarray}
where the NC field strengths $\widehat{\mathcal{F}}_{MN}$
including an instanton background are given by Eqs. \eq{f-fluc44}-\eq{conn-fluc66}.
Using the result \eq{gp-matrix}, one can calculate the fermionic term, $G \Theta \Upsilon \mathfrak{P} B
= - \kappa^2 B g^{-1} \Upsilon \mathfrak{P} B$, which takes the form
\begin{eqnarray}\label{fermi-term}
  - i \kappa^2 (Bg^{-1})_M^{~~P} \bigl( \delta^Q_P + \frac{i \kappa}{4} \overline{\psi}
  \Gamma^Q \partial_P \psi \bigr) \overline{\psi} \Gamma_Q D_N \psi
  &\equiv& - \kappa^2 (Bg^{-1})_M^{~~P} \widehat{\Upsilon}_{PN} \nonumber \\
  &\approx & - i \kappa \overline{\psi} \mathbf{\Gamma}_M D_N \psi + \mathcal{O}(\kappa^2)
\end{eqnarray}
where $\mathbf{\Gamma}_M \equiv \kappa B_{MN} g^{NP} \Gamma_P$ obey the Dirac
algebra $\{ \mathbf{\Gamma}_M, \mathbf{\Gamma}_N \} = 2G_{MN}$ and
\begin{equation}\label{fermi-covd}
    D_N \psi = \partial \psi/\partial Y^N + \{ \widehat{A}_N + \widehat{a}_N, \psi \}_\Theta.
\end{equation}
In the end, we get the supersymmetric version of Eqs. \eq{dbi-idif} and \eq{dbi-idin}:
\begin{eqnarray} \label{susy-dbi-idif}
&&\frac{1}{g_s} \int d^{10} X \sqrt{-\det \bigl( g + \kappa (\mathfrak{F} + \Upsilon) \bigr)}
 = \frac{1}{g_s} \int d^{10} Z \sqrt{-\det \bigl(\mathcal{G}  + \kappa (B + \widetilde{f} + \xi) \bigr)} \\
 \label{susy-dbi-idin}
 && \hspace{3cm} = \frac{1}{G_s} \int d^{10} Y \sqrt{-\det \bigl(G  + \kappa (\widehat{\mathcal{F}}
 + \Phi ) - \kappa^2 Bg^{-1} \widehat{\Upsilon} \bigr) }.
\end{eqnarray}

Let us redefine the fermion field, $\Psi \equiv (\kappa T_9)^{\frac{1}{2}} \psi$, and
use the approximation \eq{fermi-term} to take the expansion like Eq. \eq{det-exp}.
With this normalization, we correctly reproduce the action \eq{10dsym-action} at leading orders.
As before, we consider the limit $\Theta^{MN} \to (\zeta^{\mu\nu} = 0, \theta^{ab} \neq 0)$.
Then it is easy to see that, at nontrivial leading orders,
Eq. \eq{susy-dbi-idin} reproduces the 10-dimensional $\mathcal{N}=1$ supersymmetric NC $U(1)$
gauge theory \eq{10dsym-action} in the instanton background \eq{inst-vacuum}.
As we demonstrated in section 2, the action \eq{10dsym-action} is equivalent to the 4-dimensional
$\mathcal{N} = 4$ superconformal field theory on the Coulomb branch. And we argued in this section that
fluctuations in $AdS_5 \times \mathbb{S}^5$ background geometry are described by the
10-dimensional $\mathcal{N}=1$ supersymmetric NC $U(1)$ gauge theory in the background of NC
Hermitian $U(1)$ instantons obeying Eq. \eq{hym-eq}.
According to our construction, we thus declare that the RHS of Eq. \eq{susy-dbi-idif} has to describe
the fluctuations in $AdS_5 \times \mathbb{S}^5$ geometry.
Therefore we expect that the supersymmetric HEA for the $\mathcal{N} = 4$ superconformal field theory
on the Coulomb branch would be derived from a dimensional reduction of the RHS of Eq. \eq{susy-dbi-idif}
similar to Eq. \eq{hea-red}.

Before proceeding further, let us first address some subtle issues regarding to the equivalence in
Eqs. \eq{susy-dbi-idif} and \eq{susy-dbi-idin}. The first one is that an interpretation for
the factor $\bigl( \delta^Q_P + \frac{i \kappa}{4} \overline{\psi} \Gamma^Q \partial_P \psi \bigr)$
in $\widehat{\Upsilon}_{PN}$ is not clear from the point of view of NC $U(1)$ gauge theory.
Note that $\partial_P \psi = \partial \psi/\partial X^P$ and the Darboux transformations did not
touch the factor. Hence this factor behaves like a background part induced from the backreaction of fermions
at higher orders. Therefore a plausible picture from the viewpoint of NC $U(1)$ gauge fields is
to interpret this factor as vielbeins $\mathfrak{E}_M^A = \bigl( \delta^A_M
- \frac{i \kappa}{4} \overline{\psi} \Gamma^A \partial_M \psi \bigr)$ with an effective
metric $\mathfrak{G}_{MN} = \mathfrak{E}_M^A \mathfrak{E}_N^B g_{AB}$ and write
\begin{equation}\label{interp-dirac}
\kappa^2 (Bg^{-1})_M^{~~P} \widehat{\Upsilon}_{PN} = i \kappa \overline{\psi} \mathfrak{T}_M D_N \psi
\end{equation}
where
\begin{equation}\label{gamma-t}
    \mathfrak{T}_M \equiv \kappa B_{MN} g^{NP} \mathfrak{E}_P^A \Gamma_A.
\end{equation}
Then the gamma matrices $\mathfrak{T}_M$ satisfy the Dirac algebra
\begin{equation}\label{mod-diracalg}
    \{ \mathfrak{T}_M, \mathfrak{T}_N \} = - 2\kappa^2 (Bg^{-1}\mathfrak{G}g^{-1} B)_{MN} \equiv
    2 \mathbb{G}_{MN}.
\end{equation}
Of course, if we ignore the backreaction from the fermions, we recover the previous Dirac
term \eq{fermi-term} in flat spacetime. Another issue is how to glue local Darboux charts now
involved with fermions as well as bosons. We argued before that the global metric \eq{global-metric}
can be constructed via the globalization in terms of the gluing of local Darboux charts described
by Eqs. \eq{glue-g} and \eq{glue-d}. Or the local frames in the metric \eq{tr-abelif} are replaced
by global vielbeins \cite{mypaper}:
\begin{equation}\label{global-frame}
    \frac{\partial X^A}{\partial Z^M} \to E_M^A.
\end{equation}
Then the gamma matrices in Eq. \eq{susy-rep2} will also be replaced by $\Gamma_M
\equiv E^A_M \Gamma_A$ and $\Gamma^M \equiv E_A^M \Gamma^A$.\footnote{They should not be confused with
the gamma matrices in Eq. \eq{susy-rep1} which are defined on the flat spacetime $\mathbb{R}^{9,1}$
while those in Eq. \eq{susy-rep2} are now defined on a curved spacetime.}
Now it is also necessary to glue the fermions defined on local Darboux patches by local
Lorentz transformations
\begin{equation}\label{lorentz}
    \psi^{(j)} = S_{(ji)} \psi^{(i)}
\end{equation}
acting on fermions on an intersection $U_i \cap U_j$. As usual, we introduce a spin connection
$\omega_M = \frac{1}{2} \omega_{M AB} \Gamma^{AB}$ to covariantize the local gluing \eq{lorentz}.
This means that the fermionic terms in Eq. \eq{susy-rep2} are now given by
\begin{equation}\label{spin-conn}
\xi_{MN}  \to
i \overline{\psi} E_M^A \Gamma_A \nabla_N \psi - \frac{\kappa}{4} \overline{\psi} \Gamma^A \nabla_M
    \psi \overline{\psi} \Gamma_A \nabla_N \psi,
\end{equation}
where the covariant derivative is defined by
\begin{equation}\label{spin-cov}
 \nabla_M \psi = (\partial_M + \omega_M )\psi.
\end{equation}
The spin connections $\omega_M$ are determined by the metric \eq{asd2s5}.

Therefore the block matrix \eq{bdi-10metric} for the supersymmetric case is replaced by
\begin{equation}\label{susybdi}
\mathcal{G}_{MN} + \kappa \bigl(B + \widetilde{f} + \xi \bigr)_{MN} \approx \left(
  \begin{array}{cc}
    \lambda^2 \eta_{\mu\nu} + \kappa (f_{\mu\nu} + \xi_{\mu\nu}) & \kappa (\partial_\mu \varphi_a + \xi_{\mu a}) \\
    - \kappa (\partial_\mu \varphi_a - \xi_{a \mu}) & \lambda^2 \delta_{ab} + \kappa (B_{ab} + \xi_{ab}) \\
  \end{array}
\right).
\end{equation}
Since we are interested in the HEA of the four-dimensional supersymmetric gauge theory defined
on the boundary of $AdS_5$ space, the dimensional reduction similar to Eq. \eq{lowel-u1} was adopted too
for fermionic excitations, i.e.,
\begin{equation} \label{xi4}
\begin{array}{ll}
  \xi_{\mu\nu} = i \overline{\psi} \Gamma_\mu \nabla_\nu \psi,
  \qquad \xi_{ab} = i \overline{\psi} \Gamma_{a} \omega_{b} \psi, \\
 \xi_{\mu a} = i \overline{\psi} \Gamma_\mu \omega_a \psi,
 \qquad \; \xi_{a \mu} = i \overline{\psi} \Gamma_a \nabla_\mu \psi,
\end{array}
\end{equation}
where $\Gamma_M = E^A_M \Gamma_A$ and we ignored the quartic term in Eq. \eq{spin-conn}.
In order to get a four-dimensional picture after the dimensional reduction \eq{hea-red},
it is convenient to decompose the 16 components of the Majorana-Weyl spinor $\psi$ into
the four Majorana-Weyl gauginos $\lambda^i \; (i=1, \cdots, 4)$ as follows
\begin{eqnarray} \label{n=4spinors}
 && \psi = \left(
                                            \begin{array}{c}
                                              P_+ \lambda^i \\
                                              P_- \widetilde{\lambda}_i  \\
                                            \end{array}
                                          \right)
  \quad  \mathrm{with} \;  P_\pm = \frac{1}{2}
(I_4 \pm \gamma_5) \; \mathrm{and} \; \widetilde{\lambda}_i = - C \overline{\lambda}^{iT}, \nonumber\\
&& \Gamma^A =(\gamma^{\hat{\mu}} \otimes I_8, \gamma_5 \otimes \gamma^{\hat{a}}), \qquad
\Gamma_{11} = \gamma_5 \otimes I_8,
\end{eqnarray}
where $C$ is the four-dimensional charge conjugation operator and the hat is used to indicate
tangent space indices. We take the four- and six-dimensional
Dirac matrices in the chiral representation
\begin{eqnarray} \label{4gamma-matrix}
&& \gamma^{\hat{\mu}} = \left(
                        \begin{array}{cc}
                          0 & i \sigma^{\hat{\mu}} \\
                          -i \overline{\sigma}^{\hat{\mu}} & 0 \\
                        \end{array}
                      \right), \qquad \sigma^{\hat{\mu}} = (I_2, \vec{\sigma})
                      = (\sigma^{\hat{\mu}})_{\alpha\dot{\beta}}, \quad
                      \overline{\sigma}^{\hat{\mu}} = (-I_2, \vec{\sigma})= (\overline{\sigma}^{\hat{\mu}})^{\dot{\alpha}\beta}, \\
\label{6gamma-matrix}
&& \gamma^{\hat{a}} = \left(
                        \begin{array}{cc}
                          0 & \Sigma^{\hat{a}} \\
                          \overline{\Sigma}^{\hat{a}} & 0 \\
                        \end{array}
                      \right), \qquad \Sigma^{\hat{a}} = (\vec{\eta}, i \vec{\overline{\eta}}) = \Sigma^{{\hat{a}},ij},
                      \quad \overline{\Sigma}^{\hat{a}}
= (\Sigma^{\hat{a}})^\dagger = (-\vec{\eta}, i \vec{\overline{\eta}}) = \overline{\Sigma}^{\hat{a}}_{ij},
\end{eqnarray}
where $\vec{\sigma}$ are Pauli matrices and the $4 \times 4$ matrices
$(\vec{\eta}, \vec{\overline{\eta}})$ are self-dual and anti-self-dual 't Hooft symbols.
Then the fermion bilinear terms in Eq. (\ref{xi4}) read as
\begin{equation} \label{fbil4}
\begin{array}{l}
  \xi_{\mu\nu} = i v^{-1} \big( \overline{\lambda}_i \overline{\sigma}_{\hat{\mu}} \nabla_\nu \lambda^i
  - \lambda^i \sigma_{\hat{\mu}} \nabla_\nu \overline{\lambda}_i \big), \\
  \xi_{ab} = \partial_c v^{-1} \big( \overline{\lambda} \Sigma_{\hat{a}} \overline{\Sigma}_{\hat{b}\hat{c}} \overline{\lambda} - \lambda \overline{\Sigma}_{\hat{a}} \Sigma_{\hat{b}\hat{c}} \lambda \big), \\
 \xi_{\mu a} = 2i \partial_b v^{-1} \big( \overline{\lambda} \overline{\sigma}_{\hat{\mu}}
  \Sigma_{\hat{a}\hat{b}} \lambda \big), \\
 \xi_{a \mu} = v^{-1} \big( \overline{\lambda} \Sigma_{\hat{a}} \nabla_\mu \overline{\lambda}
 - \lambda \overline{\Sigma}_{\hat{a}} \nabla_\mu \lambda \big),
\end{array}
\end{equation}
where
\begin{equation}\label{sigma-lg}
 \overline{\Sigma}^{\hat{a}\hat{b}} \equiv \frac{1}{2} \big( \overline{\Sigma}^{\hat{a}} \Sigma^{\hat{b}} - \overline{\Sigma}^{\hat{b}} \Sigma^{\hat{a}} \big),  \qquad
 \Sigma^{\hat{a}\hat{b}} \equiv \frac{1}{2} \big(  \Sigma^{\hat{a}} \overline{\Sigma}^{\hat{b}} -
 \Sigma^{\hat{b}} \overline{\Sigma}^{\hat{a}} \big)
\end{equation}
and the spin connection for the background geometry (\ref{10vacgeo}) is given by
\begin{equation}\label{spin-back}
 \omega_\mu = - \Gamma^{\hat{\mu}\hat{a}}\partial_a \ln v, \qquad
 \omega_a = - \Gamma^{\hat{a}\hat{b}} \partial_b \ln v.
\end{equation}
Since we are considering the HEA of the four-dimensional supersymmetric gauge theory
defined on the boundary of the $AdS_5$ space where $v \to \infty$ and so the $\mathbb{S}^5$
shrinks to a point, we can ignore $\xi_{ab}$ and $\xi_{\mu a}$ in Eq. (\ref{fbil4})
as well as the spin connections $\omega_M \to 0$.

After applying the formula \eq{det-block} to the matrix \eq{susybdi}, it is straightforward to yield
the supersymmetric completion of the bosonic HEA obtained in Eq. \eq{hea-red} and it is given by
\begin{eqnarray} \label{prod-sdet}
&& \sqrt{- \det \bigl(\mathcal{G}  + \kappa (B + \widetilde{f} + \xi) \bigr)} \nonumber \\
&=& \sqrt{\det (\lambda^2 + \kappa B )} \sqrt{- \det \Bigl( \lambda^2 \eta_{\mu\nu}
+ \kappa (f_{\mu\nu} + \xi_{\mu\nu})
+ \kappa^2 \partial_\mu \varphi_a \Bigl(\frac{1}{\lambda^2 + \kappa B} \Bigr)^{ab}
(\partial_\nu \varphi_b - \xi_{b\nu}) \Bigr) } \nonumber \\
  &\approx& \lambda^6 \sqrt{- \det \Bigl( h_{\mu\nu} + \kappa (f_{\mu\nu} + \xi_{\mu\nu}
 - v^{-2} \partial_\mu \phi^a \xi_{a\nu}) \Bigr) }.
\end{eqnarray}
One may drop the last term since it is of $\mathcal{O} (v^{-3})$.
As the bosonic case (\ref{hea-red}), the 10-dimensional supersymmetric DBI action (\ref{susy-dbi-idif})
in $AdS_5 \times \mathbb{S}^5$ geometry is thus reduced to a 4-dimensional supersymmetric DBI
action given by
\begin{eqnarray}\label{shea-red}
&& - T_{D9} \int d^{10} Z  \sqrt{- \det_{10} \bigl(\mathcal{G}  + \kappa (B + \widetilde{f} + \xi) \bigr)}
\nonumber \\
&=& \Bigl( \frac{g_s N}{4\pi}\Bigr)^{\frac{3}{2}} L(\epsilon, R)
 \left[ - T_{D3} \int_W d^4 x \sqrt{- \det_4 \big(h_{\mu\nu}  + \kappa (f_{\mu\nu} + \xi_{\mu\nu}
 - v^{-2} \partial_\mu \phi^a \xi_{a\nu}) \big)} \right].
\end{eqnarray}
If the quartic term in Eq. (\ref{spin-conn}) is included, it contributes an extra term
given by $\frac{\kappa^2 v^2}{4} (\xi_{\lambda\mu} {\xi^\lambda}_\nu + \xi_{a\mu} {\xi^a}_\nu)$
inside the determinant. Since the metric (\ref{metric-sch}) becomes flat when $v=1$,
the result in this case should be equal to the action of a supersymmetric D3-brane.
One can see that the action (\ref{shea-red}) is actually the case.
See the equation (88) in Ref. \cite{sdbi12}. According to the identity (\ref{susy-dbi-idif}),
the LHS of Eq. (\ref{shea-red}) is equal to the world-volume
action of a BPS D9-brane of type IIB string theory after fixing
the $\kappa$-symmetry, which is invariant under the supersymmetry transformations given
by Eqs. (90) and (91) in Ref. \cite{sdbi12}. Since Eq. (\ref{susy-dbi-idif}) is a mathematical identity,
the action on the LHS of Eq. (\ref{shea-red}) will also be supersymmetric.
Its supersymmetry transformations basically take the form replacing the ordinary derivatives
in Eqs. (90) and (91) in Ref. \cite{sdbi12} by covariant derivatives on the $AdS_5 \times \mathbb{S}^5$ space.
But an explicit check of supersymmetry is somewhat lengthy though straightforward.
Its detailed exposition from the perspective of HEA deserves to pursue a separate work,
which will be reported elsewhere.
Note that, after the gauge fixing, $\psi_1 = 0$, for the $\kappa$-symmetry, the Wess-Zumino term
for the supersymmetric case is the same as the bosonic one \eq{scs2} \cite{sdbi12}.
The final result can be interpreted as the worldvolume action of a supersymmetric probe D3-brane
in the $AdS_5 \times \mathbb{S}^5$ background geometry.
According to the conjecture in Ref. \cite{schwarz}, it can be reinterpreted as
the HEA of four-dimensional $\mathcal{N}=4$ superconformal field theory on the Coulomb branch.
We emphasize that we directly derived the HEA from the four-dimensional $\mathcal{N}=4$
superconformal field theory on the Coulomb branch defined by the NC space \eq{n=4moyal}.

\section{Discussion}

We want to emphasize that NC spacetime should be regarded as
a more fundamental concept from which classical spacetime should be derived as
quantum mechanics is a more fundamental theory and the classical
phenomena are emergent from quantum physics.
Then the NC spacetime requires us to take a radical departure from the 20th century physics.
First of all, it introduces a new kind of duality, known as the gauge/gravity duality,
as formalized by the identity \eq{rexp-dbi}.
But we have to recall that quantum mechanics has already illustrated such kind of novel duality
where the NC phase space obeying the commutation relation $[x^i, p_j] = i\hbar \delta^i_j$ is
responsible for the so-called wave-particle duality. Remarkably there exists
a novel form of the equivalence principle stating that
the electromagnetic force can always be eliminated by a local coordinate
transformation as far as spacetime admits a symplectic structure.
The novel equivalence principle is nothing but the famous mathematical theorem known as
the Darboux theorem or the Moser lemma in symplectic geometry \cite{sg-book1,sg-book2}.
It proves the equivalence principle for the gravitational force in the context of emergent gravity.
Therefore we may conclude \cite{mypaper,hsy-jhep09} that the NC nature of spacetime is the origin
of the gauge/gravity duality and the first principle for the duality is the equivalence principle
for the electromagnetic force.

The AdS/CFT correspondence \cite{ads-cft1,ads-cft2,ads-cft3} is a well-tested gauge/gravity duality
and a typical example of emergent gravity and emergent space.
But we do not understand yet why the duality should work.
We argued that the AdS/CFT correspondence is a particular example of emergent gravity
from NC $U(1)$ gauge fields and the duality between large $N$ gauge fields and
a higher-dimensional gravity is simply a consequence of the novel equivalence principle
for the electromagnetic force. We note \cite{mypaper,hsy-jhep09} that the emergent gravity
from NC $U(1)$ gauge fields is an inevitable conclusion as far as spacetime admits a symplectic structure,
in other words, a microscopic spacetime becomes NC.
Moreover the emergent gravity is much more general than the AdS/CFT correspondence
because it holds for general background spacetimes as exemplified by the identity \eq{dbi-ccc}.
Therefore we believe that the emergent gravity from NC gauge fields provides a lucid avenue
to understand the gauge/gravity duality or large $N$ duality.

For example, it is interesting to notice that the transformation \eq{rexp-dbi} between NC $U(1)$
gauge fields and an emergent gravitational metric holds even locally.
Thus one may imagine an (infinitesimal) open patch $U$ where the field strength $F_U$ of
fluctuating $U(1)$ gauge fields has a maximal rank such that $(U, F_U)$ is a symplectic
Darboux chart. Then one can apply the Darboux theorem on the local patch
to transform the local $U(1)$ gauge fields into a corresponding local spacetime geometry supported on $U$.
But this local geometry is unfledged yet to be materialized into a classical spacetime geometry.
Hence this kind of immature geometry describes a bubbling geometry or spacetime foams which
intrinsically correspond to a quantum geometry. Even we may consider fluctuating $U(1)$ gauge fields
on a local patch $U$ whose field strengths $F_U$ do not support the maximal rank.
The dimension of emergent bubbling geometry will be determined by the rank of $F_U$ on $U$.
This implies that the dimension of quantum geometries is not fixed but fluctuates.
This picture is in a sense a well-known folklore in quantum gravity.

Then one may raise a question why NC spacetime reproduces all the results in string theory.
The connection between string theory and symplectic geometry becomes most manifest by
the Gromov's $J$-holomorphic curves. See section 7 in Ref. \cite{mypaper} for this discussion.
The $J$-holomorphic curve for a given symplectic structure is nothing but the minimal worldsheet
in string theory embedded in a target spacetime. Moreover $\alpha'$-corrections in string theory
correspond to derivative corrections in NC gauge theory. In this sense the string theory can be
regarded as a stringy realization of symplectic geometry or more generally Poisson geometry.
But the NC spacetime provides a more elegant framework for the background indepedent formulation
of quantum gravity in terms of matrix models \cite{hsy-jhep09,hsy-jpcs12} which is still elusive
in string theory.

We showed that the worldvolume effective action of a supersymmetric probe D3-brane in $AdS_5
\times \mathbb{S}^5$ geometry can be directly derived from the four-dimensional $\mathcal{N} = 4$
supersymmetric Yang-Mills theory on the Coulomb branch defined by the NC space \eq{extra-nc2n}.
Since our result, for example, described by the identity \eq{dbi-ccc} should be true for general
$U(1)$ gauge fields in an arbitrary background geometry, the remaining problem is to identify
a corresponding dual (super)gravity whose solution coincides with the emergent
metric $\mathfrak{G}_{MN}$. One may use the method in Refs. \cite{dbisugra1,dbisugra2}
to attack this problem. See also \cite{dbisugra3}. It was shown there that the worldvolume
effective action of a probe D3-brane is a solution to the Hamilton-Jacobi equation of
type IIB supergravity defined by the ADM formalism adopting the radial coordinate as time
for type IIB supergravity reduced on $\mathbb{S}^5$. In particular the radial time corresponds to
the vev of the Higgs field in the dual Yang-Mills theory as our case.
It will be interesting to find the relation between the DBI action obtained in
Refs. \cite{dbisugra1,dbisugra2} and the HEA derived in this paper.
Also there are several works \cite{cfs-p4,hea-col1,hea-col2,hea-col3,ferrari} to address
the relation of the HEA with the low-energy effective actions of $\mathcal{N}=4$
super Yang-Mills theory on the Coulomb branch. Thus it may be a vital project to understand
any relation between our approach based on the Coulomb branch defined by the NC space
and other approaches for the HEA cited above.

Recently there have been some developments \cite{asakawa-gcg,schupp-gcg} that describe D-branes
in the framework of generalized geometry. A D-brane including fluctuations in a static gauge
is identified with a leaf of foliations generated by the Dirac structure of a generalized tangent
bundle and the scalar fields and vector fields on the D-brane are unified as
a generalized connection \cite{asakawa-gcg}. It was also argued in \cite{schupp-gcg} that
the equivalence between commutative and NC DBI actions is naturally encoded in the generalized geometry
of D-branes. In particular, when considering a D-brane as a symplectic leaf of the Poisson structure,
describing the noncommutativity, the SW map is naturally interpreted in terms of the corresponding
Dirac structure. Thus NC gauge theories can be naturally interpreted within the generalized geometry.
Since the Darboux transformation relating the deformation of a symplectic structure
with diffeomorphism symmetry is one of the pillars for emergent gravity, we think that the emergent
gravity from NC gauge fields can be formulated in a natural way within the framework of generalized geometry.
It will be interesting to inquire further into this idea.

\section*{Acknowledgments}
The author thanks Hikaru Kawai and Shinji Shimasaki for warm hospitality and helpful discussions
during his visit to Kyoto University where a part of the work was done.
This work was supported by the National Research Foundation of Korea (NRF) grant funded
by the Korea government (MOE) (No. 2011-0010597).
This work was also supported by the National Research Foundation of Korea (NRF) grant funded
by the Korea government (MSIP) through the Center for Quantum Spacetime (CQUeST) of Sogang University
with grant number 2005-0049409.

\appendix

\section{K\"ahler manifolds from $U(1)$ gauge fields}

In this appendix we will illustrate how to determine four- and six-dimensional K\"ahler metrics from
$U(1)$ gauge fields by solving the identities (\ref{dbi-idc}) and (\ref{dbi-idn}) between DBI actions.
For this purpose, let us introduce $d=2n$-dimensional complex coordinates
\begin{equation}\label{comp-coord}
    z^i = x^{2i-1} + i x^{2i}, \qquad \overline{z}^i = x^{2i-1} - i x^{2i}, \qquad i=1, \cdots, n
\end{equation}
and corresponding complex $U(1)$ gauge fields
\begin{equation}\label{comp-u1g}
    A_i = \frac{1}{2} \bigl( A_{2i-1} - i A_{2i} \bigr), \qquad \overline{A}_{\bar{i}}
    = \frac{1}{2} \bigl( A_{2i-1} + i A_{2i} \bigr).
\end{equation}
Then the field strengths of $(2,0)$ and $(1,1)$ parts are, respectively, given by
\begin{eqnarray} \label{fs-20}
&& F_{ij} = \frac{1}{4} \bigl( F_{2i-1, 2j-1} -  F_{2i, 2j} \bigr) - \frac{i}{4}
\bigl( F_{2i-1, 2j} +  F_{2i, 2j-1} \bigr), \\
\label{fs-11}
&& F_{i\overline{j}} = \frac{1}{4} \bigl( F_{2i-1, 2j-1} +  F_{2i, 2j} \bigr) + \frac{i}{4}
\bigl( F_{2i-1, 2j} -  F_{2i, 2j-1} \bigr).
\end{eqnarray}
If $U(1)$ gauge fields in Eq. (\ref{comp-u1g}) are the connection of a holomorphic vector bundle,
i.e., $F_{ij}=F_{\overline{i}\overline{j}} = 0$, Eq. (\ref{fs-20}) leads to the following relations
\begin{equation}\label{holo-vb}
 F_{2i-1, 2j-1} =  F_{2i, 2j}, \qquad     F_{2i-1, 2j} = -  F_{2i, 2j-1}, \qquad i, j = 1, \cdots, n.
\end{equation}
The connections of a holomorphic line bundle can be obtained by solving the condition $F_{ij}=F_{\overline{i}\overline{j}} = 0$ and they are given by
\begin{equation}\label{holo-conn}
     A_i = - i  \frac{\partial \phi(z, \overline{z})}{\partial z^i}
     := - i \partial_i \phi(z, \overline{z}), \qquad
     \overline{A}_{\bar{i}} = i \frac{\partial \phi(z, \overline{z})}{\partial \overline{z}^i}
     = i \overline{\partial}_{\bar{i}} \phi(z, \overline{z})
\end{equation}
where $\phi(z, \overline{z})$ is a real smooth function on $\mathbb{C}^n$. Then the $(1,1)$ field
strength (\ref{fs-11}) is given by
\begin{equation}\label{holofs-11}
 F_{i\overline{j}} = 2i \partial_i \overline{\partial}_{\bar{j}} \phi(z, \overline{z}).
\end{equation}
Similarly the condition for a Hermitian metric, i.e., $\mathcal{G}_{ij}
= \mathcal{G}_{\overline{i}\overline{j}} = 0$, can be solved by
\begin{equation}\label{herm-metric}
 \mathcal{G}_{2i-1, 2j-1} =  \mathcal{G}_{2i, 2j}, \qquad
 \mathcal{G}_{2i-1, 2j} = -  \mathcal{G}_{2i, 2j-1}.
\end{equation}
If we further impose the K\"ahler condition, $d\Omega = 0$, for the Hermitian metric $ds^2
= \mathcal{G}_{i\overline{j}} dz^i d \overline{z}^j$ where
$\Omega = i \mathcal{G}_{i\overline{j}} dz^i \wedge d \overline{z}^j$
is a K\"ahler form, the metric is solely determined by a K\"ahler potential $K(z, \overline{z})$ as
\begin{equation}\label{kahl-metric}
 \mathcal{G}_{i\overline{j}} = \partial_i \overline{\partial}_{\bar{j}} \bigl( 2K(z, \overline{z})
- K_0 \bigr)
\end{equation}
where $K_0 = \overline{z}^k z^k$ and our choice of K\"ahler potential is just for a later convenience.

To deduce K\"ahler metrics from $U(1)$ gauge fields obeying Eqs. (\ref{dbi-idc}) and (\ref{dbi-idn}),
let us take their local form given by
\begin{eqnarray} \label{local-idc}
 \sqrt{\det \bigl( g + \kappa \mathcal{F} \bigr)}
&=& \sqrt{\det \bigl(\mathcal{G}  + \kappa B \bigr)} \\
 \label{local-idn}
&=& \frac{g_s}{G_s} \sqrt{\det \bigl(G  + \kappa (\widehat{F} - B) \bigr)}.
\end{eqnarray}
For our case at hand, $g_{\mu\nu} = G_{\mu\nu} = \delta_{\mu\nu}, \; \mu, \nu = 1, \cdots, d=2n$
and $B_{\mu\nu} = - \frac{2}{\kappa} \mathbf{1}_n \otimes i\sigma^2$ in Eqs. (\ref{local-idc})
and (\ref{local-idn}). We will choose the same complex structure as (\ref{comp-coord}) for all
DBI densities in Eqs. (\ref{local-idc}) and (\ref{local-idn}).
In terms of complex coordinates, their nonvanishing components are given
by $g_{i\overline{j}} = G_{i\overline{j}} = \delta_{i\bar{j}}$ and $B_{i\overline{j}}
= - \frac{i}{\kappa} \delta_{i\bar{j}}$ for $i,j = 1, \cdots, n$. Thus they are K\"ahler metrics
and a K\"ahler form on $\mathbb{C}^n$, i.e.,  $g_{i\overline{j}} = G_{i\overline{j}} = \partial_i
\overline{\partial}_{\bar{j}} K_0$ and $B_{i\overline{j}} = - \frac{i}{\kappa} \partial_i
\overline{\partial}_{\bar{j}} K_0$ with $K_0 = \overline{z}^k z^k$, respectively.
However, the RHS of Eq. (\ref{local-idc}) needs some care since $\mathcal{G}_{\mu\nu} (x)$
is regarded as a nontrivial metric on a Riemannian manifold.
For this case, it is convenient to distinguish
local coordinate indices $(\mu, \nu, \cdots)$ from tangent space indices $(a, b, \cdots)$
by introducing vielbeins $E^a_\mu$, i.e., $E^a_\mu E^a_\nu = \mathcal{G}_{\mu\nu}$.
Let us split both coordinate indices into holomorphic and antiholomorphic ones:
$\mu = (\alpha, \overline{\alpha}), \; \nu = (\beta, \overline{\beta}), \;
a = (i, \overline{i}), \; b = (j, \overline{j})$.
The Hermitian condition (\ref{herm-metric}) can be solved by taking the vielbeins as
\begin{equation}\label{holo-vielbein}
    E^i_{\overline{\alpha}} = E^{\overline{i}}_{\alpha} = 0, \qquad \quad
    E_i^{\overline{\alpha}} = E_{\overline{i}}^{\alpha} = 0.
\end{equation}
Then the nonvanishing components of $B$-field in Eq. (\ref{local-idc}) are
given by $B_{i\overline{i}} =  E_i^{\alpha} E_{\overline{i}}^{\overline{\beta}} B_{\alpha\overline{\beta}}$
where $B_{\alpha\overline{\beta}} = - i \delta_{\alpha\bar{\beta}}$.

Our primary concern is to find $U(1)$ gauge fields which give rise to the K\"ahler
metric (\ref{kahl-metric}). This means that the RHS of Eq. (\ref{local-idc}) is purely of $(1,1)$-type.
Therefore, in order to satisfy Eq. (\ref{local-idc}), the $U(1)$ gauge fields on the LHS must be
connections of a holomorphic line bundle obeying $F_{ij} = F_{\overline{i}\overline{j}} = 0$.
Moreover $\mathcal{F} = \mathcal{F}_{i\overline{j}} dz^i \wedge d\overline{z}^j$ is a nondegenerate,
closed $(1,1)$-form and so a K\"ahler form, i.e.,\footnote{Note that $F_{i\overline{j}}$ alone
in Eq. (\ref{holofs-11}) cannot be a K\"ahler form because it becomes degenerate, e.g.,
at an asymptotic infinity. This is a reason why the symplectic $B$-field is necessary
to attain a K\"ahler form.}
\begin{equation}\label{kahl-u1}
\mathcal{F} = i \partial_i \overline{\partial}_{\bar{j}} \bigl(2 \phi (z, \overline{z})
- K_0 \bigr) dz^i \wedge d\overline{z}^j
\end{equation}
because $B$ is a symplectic two-form and $F$ in Eq. (\ref{holofs-11}) satisfies the Bianchi identity,
$dF=0$. By the same reasoning, we have to impose a similar condition $\widehat{F}_{ij} = \widehat{F}_{\overline{i}\overline{j}} = 0$ for symplectic $U(1)$ gauge fields in Eq. (\ref{local-idn}).
This condition is equivalent to Eq. (\ref{holo-vb}) replaced $F$ by $\widehat{F}$.
Before proceeding to particular dimensions we are interested in, let us first discuss general properties
of the above determinant equation. Suppose that $S$ and $A$ are $d \times d$ symmetric and antisymmetric
matrices, respectively. Then we have the relation
\begin{equation}\label{det-even-odd}
   P(S,A) \equiv \det (S+A) = \det (S-A) = (-1)^d \det(-S + A).
\end{equation}
This means that the polynomial $P(S,A)$ has only even powers in $A$, or equivalently,
only even (odd) powers of $S$ appear in $P(S,A)$ for $d=$ even (odd).
When $S$ is a Hermitian metric $\mathfrak{H}$ on an $n$-dimensional (i.e., $d= 2n$) complex manifold $M$,
there is a remarkable property. As we noticed above, the DBI densities in Eqs. (\ref{local-idc})
and (\ref{local-idn}) are involved only with $(1,1)$-type quantities when we restrict ourselves
to the K\"ahler metric (\ref{kahl-metric}). The polynomial $P(\mathfrak{G}, A)$ can then be written
as the form
\begin{equation}\label{det-square}
   \det (\mathfrak{G}_{\mu\nu} + A_{\mu\nu}) = |\det (\mathfrak{G}_{\alpha\overline{\beta}}
+ A_{\alpha\overline{\beta}})|^2
\end{equation}
where $\mathfrak{G}_{\alpha\overline{\beta}} + A_{\alpha\overline{\beta}}$ is an $n \times n$ complex matrix.

The proof goes as follows. Take the LHS of Eq. (\ref{det-square}) as the form, $\det (\mathfrak{G} + A)
= \det \mathfrak{G} \; \det (1 + M)$ where ${M^\mu}_\nu = \mathfrak{G}^{\mu\lambda} A_{\lambda\nu}$.
Due to the Hermiticity property of $\mathfrak{G}$ and $A$, we have the following split:
\begin{equation}\label{m-split}
    {M^\mu}_\nu = \left\{
                    \begin{array}{ll}
                      {\mathfrak{m}^\alpha}_\beta \equiv \mathfrak{G}^{\alpha\overline{\gamma}}
A_{\overline{\gamma}\beta}, & \quad \mu = \alpha, \\
                      {\overline{\mathfrak{m}}^{\overline{\alpha}}}_{\overline{\beta}}
\equiv \mathfrak{G}^{\overline{\alpha}\gamma}
A_{\gamma\overline{\beta}}, & \quad \mu = \overline{\alpha},
                    \end{array}
                  \right.
\end{equation}
where $\mathfrak{m}$ and $\overline{\mathfrak{m}}$ are now regarded as $n \times n$ matrices.
A critical step is to use the determinant formula, $\det (1+M) = \exp {\sum_{k=1}^\infty \frac{(-)^{k+1}}{k} \mathrm{Tr} M^k}$. Then the split (\ref{m-split}) induces the same split for the trace:
\begin{equation}\label{mat-split}
 \mathrm{Tr}_{2n} M^k = \mathrm{Tr}_n \mathfrak{m}^k + \mathrm{Tr}_n \overline{\mathfrak{m}}^k
\end{equation}
where the subscript in the trace denotes the size of matrix. Therefore we get the result
\begin{equation}\label{split-square}
\det (1 + M) = \det (1 + \mathfrak{m}) \det (1 + \overline{\mathfrak{m}}).
\end{equation}
Similarly the formula, $\det \mathfrak{G} = \exp \mathrm{Tr} \ln \mathfrak{G}$,
leads to the result, $\det \mathfrak{G}_{\mu\nu} = \det \mathfrak{G}_{\alpha\overline{\beta}}
\; \det \mathfrak{G}_{\overline{\alpha}\beta}$. Combining all together,
we finally get the formula  (\ref{det-square}).

There is another interesting representation of the determinant (\ref{det-even-odd}) which
was used to formulate the kappa-symmetry of supersymmetric D-branes \cite{sdbi1,sdbi12,sdbi3,sdbi4,sdbi5,sdbi6}.
The polynomial $P(\mathfrak{G}, A)$ can be written as the form
\begin{equation}\label{pol-square}
   \det (\mathfrak{G} + A) = \rho_{\mathfrak{G}}(A)^\dagger \rho_{\mathfrak{G}} (A)
\end{equation}
where
\begin{equation}\label{rhoa}
    \rho_{\mathfrak{G}}(A) = \sum_{l=0}^{[\frac{d}{2}]} \frac{1}{2^l l! (d-2l)!} A_{\mu_1 \mu_2} \cdots
    A_{\mu_{2l-1} \mu_{2l}} \gamma_{\mu_{2l+1} \cdots \mu_d} \varepsilon^{\mu_1 \cdots \mu_d}.
\end{equation}
Here $\gamma$-matrices on $M$ are defined as usual as $\gamma_\mu = E^a_\mu \gamma_a$
and the $\gamma$-matrices $\gamma_a$ obey the Dirac algebra $\{\gamma_a, \gamma_b \} = 2 \delta_{ab}$
on flat space. For the proof of Eq. \eq{pol-square}, see, in particular, Appendix A in Ref. \cite{sdbi12}
and Appendix B in Ref. \cite{sdbi3}.
See also \cite{koerber} (eq. (2.18)). It is convenient to introduce the skew-exponential
function \cite{sdbi6} (the usual exponential function with completely skew-symmetrized indices
of gamma matrices at every order in the expansion)
\begin{equation}\label{se-def}
    \mathrm{se}^{-\mathbb{A}} = \sum_{l=0}^{[\frac{d}{2}]} \frac{(-1)^l}{2^l l!}
    \gamma^{\mu_1 \cdots \mu_{2l}}
    A_{\mu_1 \mu_2} \cdots A_{\mu_{2l-1} \mu_{2l}}
\end{equation}
to rewrite $\rho_{\mathfrak{G}}(A)$ as
\begin{equation}\label{se-rewrite}
    \rho_{\mathfrak{G}}(A) = \mathrm{se}^{-\mathbb{A}} \Gamma_{\mathfrak{G}}
\end{equation}
where $\mathbb{A} \equiv \frac{1}{2} \gamma^{\mu\nu} A_{\mu\nu}$ and
\begin{equation}\label{vol-gamma}
 \Gamma_{\mathfrak{G}} = \varepsilon^{\mu_1 \cdots \mu_{d}} \gamma_{\mu_1 \cdots \mu_{d}}
 = (-i)^{\frac{d(d-1)}{2}} \sqrt{\det \mathfrak{G}} \gamma_{d+1}.
\end{equation}
Using the formula (\ref{se-rewrite}), we get the skew-exponentials for each DBI density:
\begin{eqnarray}\label{equiv-square1}
   && \rho_{g}(\mathcal{F}) = (-i)^{\frac{d(d-1)}{2}} \mathrm{se}^{-\gamma^{i\overline{j}} \mathcal{F}_{i\overline{j}}} \gamma_{d+1}, \\
\label{equiv-square2}
   && \rho_{\mathcal{G}}(B)
    = (-i)^{\frac{d(d-1)}{2}} \sqrt{\det \mathcal{G}} \; \mathrm{se}^{-\gamma^{\alpha\overline{\alpha}}
    B_{\alpha\overline{\alpha}}} \gamma_{d+1}, \\
 \label{equiv-square3}
  && \rho_G(\widehat{\mathcal{F}}) = (-i)^{\frac{d(d-1)}{2}} \mathrm{se}^{-\gamma^{i\overline{j}} \widehat{\mathcal{F}}_{i\overline{j}}} \gamma_{d+1},
\end{eqnarray}
where $\widehat{\mathcal{F}} \equiv \widehat{F} - B$ and $\gamma_{d+1}^2 = 1$.
We set $\kappa = 1$ for convenience.

Note that, using the results (\ref{kahl-metric}) and (\ref{kahl-u1}), we get the expression
\begin{eqnarray} \label{kp-gauge}
 && g_{i\overline{j}} + \mathcal{F}_{i\overline{j}} =
i \partial_i \overline{\partial}_{\bar{j}} \bigl(2 \phi - K_0 - iK_0 \bigr), \\
 &&  \mathcal{G}_{i\overline{j}} + B_{i\overline{j}} =
\partial_i \overline{\partial}_{\bar{j}} \bigl( 2K - K_0 - iK_0 \bigr)
\end{eqnarray}
where we did not discriminate curved and flat space indices because it is no more necessary.
Now, using the relation (\ref{det-square}), we can phrase the equivalence (\ref{local-idc})
in terms of K\"ahler potentials (up to holomorphic gauge transformations):
\begin{equation}\label{equiv-kahler}
\phi (z, \overline{z}) = K (z, \overline{z}).
\end{equation}
The real function $\phi (z, \overline{z})$ and so the K\"ahler potential $K (z, \overline{z})$
will be determined by solving the equations of motion of either commutative or NC $U(1)$ gauge fields.
We remark that the relation (\ref{equiv-kahler}) is completely consistent with that
in Ref. \cite{hsy-prl} (see Eqs. (30) and (31)) for the equivalence
between hyper-K\"ahler manifolds and symplectic $U(1)$ instantons. (See also \cite{hsy-epl}.)
Therefore the relation (\ref{equiv-kahler}) generalizes the one in \cite{hsy-prl,hsy-epl}
to general $2n$-dimensional K\"ahler manifolds. Recall that the Ricci tensor and the Ricci-form for
a $2n$-dimensional K\"ahler manifold are given by
\begin{equation}\label{n-ricci12}
    R_{i\overline{j}} = - \frac{\partial^2 \ln \det \mathfrak{G}_{k\overline{l}}}{\partial z^i
\partial \overline{z}^j}, \qquad \rho = - i \partial \overline{\partial}
\ln \det \mathfrak{G}_{i\overline{j}},
\end{equation}
respectively. In particular, the Ricci tensor (\ref{n-ricci12}) vanishes if $\det \mathfrak{G}_{i\overline{j}}$
is constant and so the K\"ahler manifold reduces to a $2n$-dimensional Calabi-Yau manifold.
Hence we can translate the statement for K\"ahler manifolds into that for $U(1)$
gauge theory and vice versa using the relation (\ref{equiv-kahler}).
For example, one may wonder what is the gauge theory object that gives rise to the $2n$-dimensional
Calabi-Yau manifold. It was verified in \cite{hsy-prl,hsy-epl} for the four-dimensional case that
it is the commutative limit of NC $U(1)$ instantons \cite{4nc-instanton}.
Later it was conjectured in \cite{mypaper} that Calabi-Yau $3$-folds arise from a semiclassical
limit of NC Hermitian $U(1)$ instantons in six dimensions.

Now we will show that the conjecture in \cite{mypaper} is true. First we will illustrate our method
with the four-dimensional case since this case was well established in \cite{hsy-prl,hsy-epl}.
Then we will generalize our approach to the six-dimensional case. Consider four-dimensional symplectic
$U(1)$ instantons as the commutative limit of NC $U(1)$ instantons \cite{4nc-instanton} obeying
the self-duality equations
\begin{equation}\label{nc-u1inst}
    \widehat{F}_{\mu\nu} = \pm \frac{1}{2} {\varepsilon_{\mu\nu}}^{\rho\sigma} \widehat{F}_{\rho\sigma}
\end{equation}
or in a compact notation
\begin{equation}\label{nc-cninst}
    P_\mp \widehat{\mathbb{F}} = 0
\end{equation}
where $P_\pm = \frac{1}{2} (1 \pm \gamma_5)$ and $\widehat{\mathbb{F}} = \frac{1}{2} \gamma^{\mu\nu}
\widehat{F}_{\mu\nu}$. In terms of complex coordinates (\ref{comp-coord}),
the self-duality equations (\ref{nc-u1inst}) can be stated as\footnote{The complex structure
in Eq. (\ref{sd-holo}) is correlated with the self-dual structure in Eq. (\ref{nc-u1inst}).
In this appendix we will fix the complex structure with the coordinates (\ref{comp-coord}).
Instead we will flip the orientation for the definition of the self-duality equations (\ref{nc-u1inst}),
e.g., $\varepsilon^{1 2 \cdots (2n) (2n-1)} = 1$ for the self-dual case and
$\varepsilon^{1 2 \cdots (2n-1) 2n} = 1$ for the anti-self-dual case.}
\begin{eqnarray} \label{sd-holo}
&&  \widehat{F}_{ij} = \widehat{F}_{\overline{i}\overline{j}} = 0, \\
\label{sd-11}
&&  \widehat{F}_{i \overline{i}} = 0.
\end{eqnarray}
In order to see what kind of condition the instanton equations (\ref{sd-holo}) and (\ref{sd-11})
impose on the K\"ahler metric $\mathcal{G}_{i\overline{j}}$, let us apply the SW map (\ref{sw-mapf})
to them. An important part is to note that $\theta^{i\overline{j}} = - i \delta^{i\bar{j}}$ or
$\theta^{2i-1, 2j} = \frac{1}{2} \delta^{ij}$ due to the relation $B_{\mu\lambda} \theta^{\lambda\nu}
= \delta^\nu_\mu$. Then it is easy to see that Eq. (\ref{sd-holo}) can be solved by $F_{ij} = F_{\overline{i}\overline{j}} = 0$ for which ${N_{\mu}}^\nu \equiv \delta^\nu_\mu
+ F_{\mu\lambda} \theta^{\lambda\nu}$ is split into holomorphic and anti-holomorphic parts like as
Eq. (\ref{m-split}). In particular, ${N_{i}}^j = \delta^j_i + F_{i\overline{k}} \theta^{\overline{k}j}
= \delta_{i\overline{j}} + i F_{i\overline{j}} = - \partial_i \overline{\partial}_{\bar{j}} (2 \phi - K_0)
= - \mathcal{G}_{i \bar{j}}$ where Eqs. (\ref{holofs-11}) and (\ref{equiv-kahler})
were used. Then we can easily solve Eq. (\ref{sd-11}):
\begin{equation}\label{sde-11}
   \widehat{F}_{i \overline{i}} = {(N^{-1})_i}^k F_{k \overline{i}} = - i {(N^{-1})_i}^k
\bigl( {N_{k}}^i - \delta_k^i \bigr) = -i (2 - \mathrm{Tr} N^{-1}) = 0.
\end{equation}
Using the relation $\mathrm{Tr} N^{-1} = \mathrm{Tr} N/\det N$, we get $\mathrm{Tr} N= 2 \, \det N$.
Motivated by this relation, we define a new matrix $\mathfrak{G}$ as $N = \frac{1}{2} (1 + \mathfrak{G})$
so that $\det \mathfrak{G}_{i\overline{j}} = 1$. In consequence the K\"ahler metric
$\mathfrak{G}_{i\overline{j}}$ is Ricci-flat because of the formula (\ref{n-ricci12}).
In other words, the four-manifold described by the metric $\mathfrak{G}_{i\overline{j}}$
is a hyper-K\"ahler manifold or a Calabi-Yau two-fold. In the end we have checked the equivalence
between symplectic $U(1)$ instantons and Calabi-Yau 2-folds in \cite{hsy-prl,hsy-epl}.\footnote{Note
that we are solving the determinant equations (\ref{local-idc}) and (\ref{local-idn}) and
so $\mathcal{G}_{i \bar{j}} = - N_{i\overline{j}}$ leads to the relation $\mathcal{G}_{\mu\nu}
= \frac{1}{2} (\delta_{\mu\nu} + \mathfrak{G}_{\mu\nu})$ according to the formula (\ref{det-square}),
which was used in \cite{hsy-prl,hsy-epl} to identify a gravitational metric $\mathfrak{G}_{\mu\nu}$
from the emergent metric $\mathcal{G}_{\mu\nu}$ determined by $U(1)$ gauge fields.}

Now we consider the six-dimensional case. The analysis is almost the same as the four-dimensional case.
We consider six-dimensional symplectic $U(1)$ instantons satisying the Hermitian
Yang-Mills equations \cite{non-inst}
\begin{equation}\label{nc-hym}
    \widehat{F}_{\mu\nu} = - \frac{1}{4} {\varepsilon_{\mu\nu}}^{\rho\sigma\alpha\beta}
    \widehat{F}_{\rho\sigma} I_{\alpha\beta}
\end{equation}
where $I = \mathbf{1}_3 \otimes i \sigma^2$ is a complex structure of $\mathbb{R}^6$. They can be written
with the complex coordinates (\ref{comp-coord}) and the result takes the same form as Eqs. (\ref{sd-holo})
and (\ref{sd-11}). The same argument shows that Eq. (\ref{sd-holo}) can be solved by $F_{ij} = F_{\overline{i}\overline{j}} = 0$ and Eq. (\ref{sd-11}) leads to the result $\widehat{F}_{i \overline{i}} = {(N^{-1})_i}^k F_{k \overline{i}} = -i (3 - \mathrm{Tr} N^{-1}) = 0$, i.e., $\mathrm{Tr} N^{-1} = 3$.
The trace of $3 \times 3$ complex matrix $N^{-1}$ is given by
\begin{equation}\label{33n}
    \det N \mathrm{Tr} N^{-1} = N_{1\overline{1}} N_{2\overline{2}} + N_{2\overline{2}} N_{3\overline{3}}
    + N_{3\overline{3}} N_{1\overline{1}} - \bigl( N_{1\overline{2}} N_{2\overline{1}} + N_{2\overline{3}} N_{3\overline{2}} + N_{3\overline{1}} N_{1\overline{3}} \bigr).
\end{equation}
By a similar reasoning to the four-dimensional case, we introduce a new metric $\mathfrak{G}$ defined by
$N = \frac{1}{3} (1 + \mathfrak{G})$. A straightforward calculation shows that $\mathrm{Tr} N^{-1} = 3$
can be written as the form
\begin{equation} \label{det-tr}
    \det \mathfrak{G} = 2 + \mathrm{Tr} \mathfrak{G}.
\end{equation}
Note that $\varphi \equiv i \mathfrak{G}_{i\overline{j}} dz^i \wedge d\overline{z}^j$ is a closed two-form
of type (1,1) and so we may assume, up to an addition of an exact two-form, that $\varphi$ is harmonic.
And the trace $\mathrm{Tr} \mathfrak{G}$ is equal to the contraction of $\varphi$ with
the K\"ahler form $\omega \equiv \frac{1}{2} I_{\mu\nu} dx^\mu \wedge dx^\nu$, i.e.,
$\mathrm{Tr} \mathfrak{G} = (\varphi, \omega)$. Since $\varphi$ is a harmonic (1,1)-form,
its trace $\mathrm{Tr} \mathfrak{G}$ is then constant \cite{besse} (see $\mathbf{2.33}$).
In consequence, the six-manifold described by the metric $\mathfrak{G}_{i\overline{j}}$
is a Ricci-flat and K\"ahler manifold, i.e., a Calabi-Yau 3-fold. Therefore we confirm
the conjecture in \cite{mypaper} for the equivalence between Hermitian $U(1)$ instantons
and Calabi-Yau 3-folds.

In order to check our conjecture for the $AdS_5 \times \mathbb{S}^5$ geometry,
it is necessary to sum up the stack of Hermitian $U(1)$ instantons obeying (\ref{nc-hym}).
This may be a challenging problem and we do not know yet how to sum up the lump of infinitely
many Hermitian $U(1)$ instantons near the origin of $\mathbb{R}^6$.
We leave this problem and an explicit construction of emergent K\"ahler metrics for future works.

\end{document}